# Fractional Laplacian Spectral Approach to Turbulence in a Dusty Plasma Monolayer

E G Kostadinova [1, 2, a)], R Banka [1], J L Padgett [1, 3], C D Liaw [1, 4], L S Matthews [1], & T W Hyde [1]

[1] CASPER and Department of Physics, Baylor University, Waco, TX, 76706, USA

[2] Physics Department, Leach Science Center, Auburn University, Auburn, AL, 36849, USA

[3] Department of Mathematical Sciences, University of Arkansas, Fayetteville, AR, 72701, USA

[4] Department of Mathematical Sciences, University of Delaware, Newark, DE 19716, USA

[a)] Author to whom correspondence should be addressed: egk0033@auburn.edu

**Abstract.** This work presents an analytical investigation of anomalous diffusion and turbulence in a dusty plasma monolayer, where energy transport across scales leads to the spontaneous formation of spatially disordered patterns. Many-body simulations of 10,000-particle dusty plasma monolayers are used to demonstrate how the global dynamics depend on the statistical properties of the dust assembly for realistic laboratory conditions. We find that disorder due to variations in the dust size distribution and charge-driven nonlocal interactions resulting in anomalous dust diffusion are key factors for the onset of instabilities. The resulting dynamics exhibit features of inertial turbulence over slightly more than half a decade of scales proportional or smaller than the Debye shielding length. These processes are examined analytically using a recently developed Fractional Laplacian Spectral (FLS) technique, which identifies the active energy channels as a function of scale, disorder concentration, and features of the nonlocal interactions. The predictions from the theoretical (spectral) analysis demonstrate agreement with the results from the many-body (kinetic) simulations, thus providing a powerful tool for the study of active turbulence.

## I. INTRODUCTION

Chaos and nonlinearity are ubiquitous in the natural world, yet the current understanding of many related phenomena is limited by the lack of solid theoretical framework. An outstanding example is the origin of turbulence in charged media. Unlike conventional fluids, where turbulent behavior is typically dominated by inertial forces, the excitation of instabilities in charged fluids is further influenced by electromagnetic forces and collective effects [1]. In addition, dynamical instabilities, such as vortices, can lead to a redistribution of charge throughout the fluid, resulting in modification of the electric field. These effects are particularly prominent in systems where inertial (heavy) particles are suspended within the flow [2], [3]. Such systems include atmospheric clouds [4]–[6], fluidized bed reactors [7]–[9], charged industrial sprays [10], and climate-related dust lifting processes [11], [12]. Thus, knowledge of the fundamental physics mechanisms underlying the coupling between charge-related phenomena and the onset of instabilities is key to understanding turbulence across a wide range of physical systems.

Investigation of such mechanisms is especially interesting at low-to-medium Reynolds numbers ($Re \lesssim 100$), where the fluid dynamics *are not* solely dominated by inertial effects. The onset of turbulence in this regime is important in the study of highly viscous liquids or media with small-scale flows, including viscoelastic fluids [13]–[16], porous materials [17]–[19], and airfoil devices [20]–[22]. The ubiquitous presence of charge inhomogeneities and inertial particle suspensions in these systems requires knowledge of the coupling between charge-mediated interactions and small-scale excitations. Another interesting question is turbulence in active systems, such as insect flights [23], [24], and stochastic self-propelled devices [25], [26], where energy driven at small scales can be transported to larger scales, causing a global instability. As these various systems are often characterized by small-scale, high-frequency dynamics, which violate the assumptions needed for a hydrodynamic approximation, investigation of the onset of turbulence in this case requires a kinetic treatment.

The complexity of phenomena affecting the onset of turbulence in charged media and/or active systems can be studied at the kinetic level employing complex (dusty) plasma as a model system. The field of complex plasmas investigates the dynamics of mesoscopic particles (or dust) suspended in weakly-ionized low temperature gas. Dust grains immersed in plasma become negatively charged and are subject to both ion drag forces and collective interactions. Dusty plasmas can self-organize into strongly coupled fluids and crystalline structures [27]–[30], which makes them ideal for the study of self-organization and stability, phase transitions, and transport phenomena. Experiments and numerical simulations using dusty liquids at low-to-medium Reynolds numbers have already been used to analyze self-excited [31]–[33] and externally-driven [34], [35] vortices, shear instabilities [36], [37], Kolmogorov flows [38], and wave turbulence [39]–[41]. More importantly, the mesoscopic particles, which are the tracers of the turbulent dynamics in a dusty plasma liquid, are directly observable on easily accessible spatial and temporal scales, which allows an investigation at the kinetic level of the connection between global instabilities and individual particle statistics.

Here we study the onset of turbulence in a dusty plasma monolayer, where energy is driven at the level of each individual particle and transported to larger scales, causing an 'inverse' cascade, as opposed to the classical Kolmogorov cascade [42]. As an inverse cascade is typical for two-dimensional (2D) turbulence, it has been observed in active matter monolayer systems, such as bacterial suspensions [43] and self-propelled camphor swimmers [44]. Dusty plasma monolayers, or quasi-2D structures, are intrinsically non-equilibrium systems, where energy is constantly driven and dissipated at small scales due to the interaction of individual dust particles with the neutral gas and plasma environment. Additionally, dust charging processes in these systems yield both nonlocal interactions and stochastic fluctuations in the dust particle spatial distribution. Thus, the investigation of self-excited instabilities in dusty plasma can provide insight into the connection between charge-driven processes and the inverse energy cascade in 2D turbulence.

This paper presents many-body (MB) simulations where dusty plasma monolayers are formed in conditions relevant to laboratory experiments. Specifically, we assume a distribution of dust particle sizes consistent with values reported by dust particle suppliers. Deviations from the mean particle size result in fluctuations of the corresponding particle charge, interparticle spacing, and interaction potential, all of which determine the dusty plasma dynamics and stability. To mimic conditions achievable in dusty plasma experiments in large electrode chambers [45], we consider 10,000-particle monolayers with radial confinement acting most strongly at the cloud exterior, as set by a tenth-order polynomial confinement force. The kinetic energy in the system is regulated by the gas pressure: collisions between dust and neutral gas particles provide both a drag force, which damps particle motion, as well as a thermal bath, which provides 'kick' excitations of various strength and direction to each dust grain.

Comparisons between stable and unstable monolayers are made by considering three pressure regimes, $p = 5\ Pa, 1\ Pa,$ and $0.1\ Pa$, which allows for simulation of both crystalline and liquid-like monolayers. At $5\ Pa$, a stable monolayer is formed with distinct hexagonal symmetry and large crystalline domains. At $1\ Pa$, both the gas drag force and the strength of kicks from the thermal bath are decreased and the dust particles are mobilized, which results in enhancement of defect lines and shrinking of the crystalline regions. As the pressure drops to $0.1\ Pa$, the crystalline domains are destroyed, and the monolayer develops an instability. In each case, energy is imported at the individual particle level by the thermal bath and transported across larger spatial scales through electrostatic scattering events, which are enhanced at low pressure due to increased particle mobility. This allows for an inverse energy cascade from small to large scales for given system parameters. The direct Kolmogorovian energy cascade from large to small scales is also possible in dusty plasmas, for example, using electron beams or optical lasers to induce a large-scale shear flow in the structure [46], [47]. This concept will be explored in our future work. The connection between 2D turbulence in dusty plasma monolayers and inertial turbulence is briefly discussed in Sec. III A.

For each set of conditions, the observed dust dynamics is compared against the predictions from the Fractional Laplacian Spectral (FLS) model, which determines the time-evolved energy state of the systems solely from knowledge of the underlying Hamiltonian structure. The spectral approach employed here (introduced in [48], [49]) identifies whether the energy spectrum of a given Hamiltonian operator includes an absolutely continuous part, which indicates the existence of transport in the form of scattering energy states. To explore the effect of random disorder and nonlocal interactions on the energy transport, the Hamiltonian used in this work is the *random fractional discrete Schrödinger operator* [50]. The potential energy term of this operator models random disorder, while nonlocality is allowed by employing a fractional Laplacian operator for the kinetic energy term. However, the spectral approach is a general mathematical tool which can be applied to any Anderson-type Hamiltonian of the form defined in [51].

Scaling between the many-body (MB) simulation and the FLS technique is performed in the following manner. In the MB simulation, the variation of dust size (and, therefore, mass and charge) and properties of the radial confinement result in fluctuations of the mean interparticle separation and interparticle potential in space. The standard deviation from the mean particle spacing is used as a measure of random disorder, which determines the corresponding potential energy term in the Hamiltonian $H$ used in the FLS calculation. The kinetic energy term in $H$ is obtained from the characteristics of the dust particle diffusion observed in the MB simulation (the plot of mean squared displacement as a function of time and the probability distribution function of particle velocities). Once a Hamiltonian for a given set of MB simulation conditions is determined, the FLS technique is used to identify the existence of transport as a function of spatial scales, solely from the properties of the corresponding energy spectrum. Thus, an agreement between the (kinetic) may-body simulations and the (spectral) FLS calculations provides a powerful modeling tool for the study of energy transport in turbulent dynamics.

In Sec. II, we introduce the main features of the many-body simulation and discuss how realistic experimental conditions are scaled as input parameters for the simulation. Structural and statistical analyses of the resulting dusty plasma monolayers are presented in Sec. III, where we also discuss scaling between the observed properties of the monolayers and the corresponding Hamiltonian structure. In Sec. IV, the FLS method is introduced and used to compute the probability for energy transport as a function of scale for each set of conditions. Discussion of the observed similarities between the kinetic simulation and the spectral model is provided in Sec. V. In Sec. VI, we summarize the results and outline directions for future research. More details of the many-body simulation are discussed in Appendix A. Detailed description of the FLS method and related mathematical proofs can be found in [50]. Appendix B includes further details on the velocity distribution functions obtained from the many-body simulations.

## II. SIMULATION OF DUSTY PLASMA MONOLAYERS

### A. Scaling of Experimental Parameters

An outstanding challenge in the experimental study of global instabilities in complex plasmas on Earth is the need to generate and sustain extended structures, consisting of many dust particles. A main motivation for the present numerical study is the recent development of dusty plasma experiments where extended monolayers, consisting of more than 10,000 particles, have been achieved. One such experimental setup is Cell 3 (Fig 1a), located in Baylor University's CASPER lab. Cell 3 is based on the GEC RF reference cell chamber [52], with several modifications to the original design required for suspending dust particles in the discharge and visualizing them using different optical systems [53]–[55]. A unique feature of Cell 3 is its large geometrical size, with an 8-inch ($\approx$ 20.3 cm) diameter lower electrode, which allows for the formation of very large dust clouds. Experiments with Cell 3 have already produced dust crystals (monolayers and multi-layered structures), with radial spread of about 7.5 in ($\approx$ 19 cm). Figure 1b shows the outer region of such a dusty plasma monolayer suspended in argon plasma at gas pressure $1.33\ Pa$. At this pressure, the structure exhibits both crystalline domains and liquid-like disordered regions, which suggests the possibility of interesting energy transport throughout the structure.

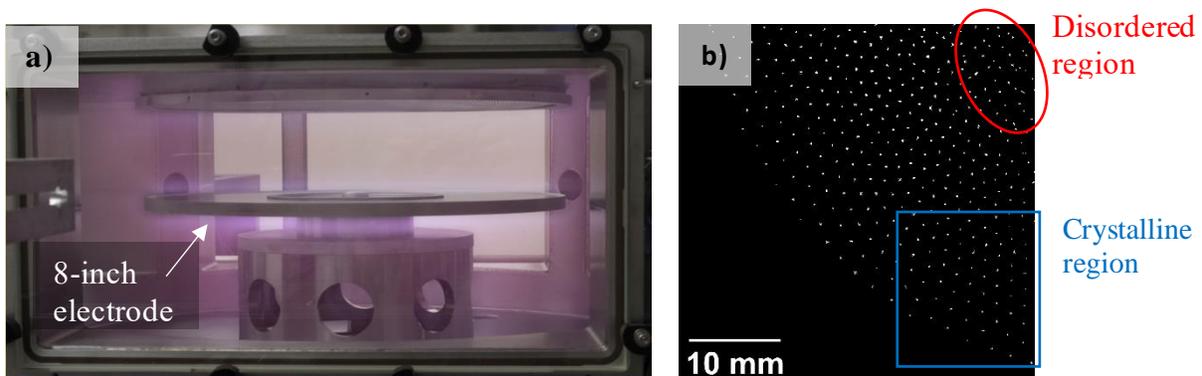

Fig. 1. a) Cell 3: a large-electrode RF cell, located at Baylor University's CASPER lab. b) Outer region of an extended dusty plasma monolayer generated in Cell 3, argon plasma, pressure $1.33\ Pa$. At this pressure, the monolayer exhibits both crystalline domains and liquid-like disordered regions.

To the best of our knowledge, dust structures of comparable diameter have been reported in only one other modified GEC RF cell within the dusty plasma community, as described in Meyer *et al.* [45]. The structures discussed in Meyer *et al.* consist of $\approx 10,000$ dust particles, forming monolayers of diameter $\approx 27 cm$. At $0.15\ Pa$, these extended monolayers were observed to exhibit wave instabilities and mode coupling, which suggests that this lower pressure is appropriate for modeling fluid dynamics phenomena, such as turbulence.

Motivated by the experimental realization of extended dusty plasma structures in large chambers, here we focus on modeling monolayers consisting of $10,000$ particles at pressures $1\ Pa$ and $0.1\ Pa$; conditions for which a liquid-like dynamics have been observed. For completeness and comparison to more ordered structures, we also consider a case where the monolayer is generated at $5\ Pa$. At this higher pressure, neutral gas drag stabilizes the dust particle motion and the resulting monolayers are expected to exhibit large-scale crystalline domains and well-defined hexagonal symmetry (e.g., see the experiments reported in [30].) Table I summarizes the experimentally relevant conditions used as inputs for the many-body simulations of dusty plasma monolayers at the three considered pressure cases.

Table I. Conditions used in the many-body simulations of dusty plasma monolayers

| Dust Parameters | | Gas / Plasma Environment | | | |
|---|---|---|---|---|---|
| Type | MF spheres | discharge | RF glow, Argon | | |
| Diameter [$\mu m$] | $9.19 \pm 0.08$ | Pressure [$Pa$] | 5 | 1 | 0.1 |
| Mass [$\times 10^{-13} kg$] | $6.1 \pm 0.2$ | $\lambda_D$ [$mm$] | 0.6 | 1 | 1 |
| $<Q_d>$ [$e^-$] | 24,648 (1, 0.1 $Pa$) <br> 18,570 (5 $Pa$) | $\Omega_h [s^{-1}]$ | $2\pi \times 0.12$ | | |
| Particle number | 10,000 | Frame rate [$fps$] | 100 | | |

In Table I, the dust charge $Q_d$ is in units of elementary charge $e^-$, $\lambda_D$ is the Debye screening length due to the electron and ions in the plasma, and $\Omega_h$ is the radial confinement frequency (discussed in the next section). For the low pressure cases, $p = 1\ Pa, 0.1\ Pa$, the selected Debye length is $\lambda_D = 1 mm$, which has been reported as a typical value at lower pressures in Argon RF glow discharge [45], [56]. In [45], the charge on the dust grains was estimated to $Q_d = 24,648 e^-$ which accounts for the reduction of charge due to interactions with ion wakefields. At $p = 5 Pa$, the higher plasma density leads to a reduced Debye length of $\lambda_D = 0.6 mm$, as discussed in [30]. Since dust charge reduction due to ion-neutral collisions is also expected, in this case the simulation employs a dust charge $Q_d = 18,570 e^-$, as suggested in [57]. In Table I, the values of $Q_d$ represent the mean charge on the dust particles. However, dust particles used in laboratory experiments are known to exhibit slight variations from the mean particle diameter, which can lead to observable differences in the monolayer dynamics, such as the bistability reported in [56]. Thus, here the dust particle diameters are selected from a normal distribution with a mean $9.19 \mu m$ and a standard deviation of $\pm 0.9\%$. This also leads to variation in the corresponding mass and charge on each dust grain.

## B. Main Features of the Many-Body Simulation

To simulate the dynamics of dusty plasma monolayers, we employ an $N$-body code optimized to include specified external potentials suitable for modeling the confinement forces and external electric fields characteristic of laboratory dusty plasma experiments. The code explicitly simulates the dynamics of the dust particles. The presence of other species in a dusty plasma (electrons, ions, and neutral atoms) is implicitly accounted for by including the assumed effect of these species on the dynamics of the dust. The code allows for specification of dust particle parameters, such as size, charge, and material density. In this work, the interaction between the dust grains and the surrounding plasma environment is simulated by choosing the appropriate Debye shielding length and form of the interaction potential. This code has been extensively used in modeling the dynamics of charged dust in astrophysical environments [58]–[62] and in GEC RF reference cells in Earth-based experiments [63]–[66]. In all cases, we use a simulation box size with 85 $cm$ sides, which was selected to mimic the $85 cm$-diameter electrode chamber used for the experiments in [45].

In all simulations, we use a 4th-order Runge-Kutta adaptive time step with minimum time step $10^{-6}$ $s$ and maximum time step $10^{-4}$ $s$. As a rule of thumb, the selected time step needs to be small enough to resolve the particle motion on a meaningful scale for the velocities involved. As discussed in the next section (see Fig. 2), for the 5 Pa and 1 Pa cases, the horizontal and vertical velocities are $\sim 10^{-5}$ $ms^{-1}$, while in the 0.1 Pa case, the vertical velocity for individual particles is as large as $\sim 10^{-2}$ $ms^{-1}$. Therefore, in all simulations, the maximum particle motion is $\sim (10^{-4}\ s) \times (10^{-5}\ ms^{-1}) = 10^{-9}\ m$ for the 0.1 Pa and 5 Pa cases, and $\sim (10^{-4}\ s) \times (10^{-2}\ ms^{-1}) = 10^{-6}\ m$ for the 0.1 Pa case. As the dust size is $\approx 10\ \mu m = 10^{-5}\ m$, the selected timestep can well resolve the dust motion.

In the many-body simulation, the equation of motion for a single dust particle with mass $m_d$ and charge $Q_d$ is given by

$$m_d \ddot{\pmb{x}} = \pmb{F}_{dd} + m_d \pmb{g} + Q_d \pmb{E} - \beta \dot{\pmb{x}} + \zeta \pmb{r}(t), \tag{1}$$

where $\pmb{F}_{dd}$ is the force between pairs of dust grains, $m_d \pmb{g}$ is the gravitational force, $Q_d \pmb{E}$ is the confining electric field force, and $\beta \dot{\pmb{x}}$ is the neutral drag force. The force between pairs of dust grains $\pmb{F}_{dd}$ is assumed to be a shielded Coulomb (Yukawa) interaction, where the screening is mainly provided by the ions and determined from the Debye length $\lambda_D$. The last term $\zeta \pmb{r}(t)$ is a thermal bath, which simulates diffusive motion of the dust grains due to random collisions with the neutrals in the gas. Here, $\zeta$ is the maximum acceleration exerted from the neutrals to each dust particle and r(t) is a random number selected from a normal distribution. Further details on the dust-dust interaction force, the axial (along the direction of gravity) electrostatic confinement force, neutral drag force, and thermal bath can be found in Appendix A.

To model the radial electrostatic confinement in large-electrode cells, the radial potential acting on the $i^{th}$ particle is a tenth-order polynomial of the form suggested in [45]

$$V_i = 0.5 m_d \left( \frac{\Omega_h^2 \rho_i^{10}}{R^8} \right), \tag{2}$$

where $\Omega_h$ is the horizontal confinement frequency, $\rho_i$ is the radial position of the $i^{th}$ particle in the $xy$-plane, and $R$ is the approximate horizontal radius of the monolayer (radial extent). Following Meyer *et al.* [45], we assume $\Omega_h = 2\pi \times 0.12\ s^{-1}$ and $R = 63\ mm$. Data for the particle positions, velocities, and accelerations are output every 0.001 s, which is selected to match a frame rate of 1000 $fps$, typically needed to resolve the motion of fast particles in laboratory dusty plasma experiments. Thus, in the following discussion, timesteps in the many-body simulation are sometimes referred to as frames.

### III. STRUCTURE AND DYNAMICS OF SIMULATED MONOLAYERS

Using experimentally relevant conditions from Table 1, simulations of extended dusty plasma monolayers were conducted for the three pressure cases, $p = 5\ Pa, 1\ Pa,$ and, $0.1\ Pa$, which are expected to yield different structure and dynamics. Figures 2abc show the dust particle positions after 30s of simulation time, which is sufficient to allow for damping of instabilities due to initial conditions. Comparison of the three pressure cases clearly demonstrates that the monolayers form a stable structure at $5\ Pa$ and $1\ Pa$, while the $0.1\ Pa$ case yields a less ordered, liquid-like structure. In each simulation, the properties of the radial confinement in equation (2) are fixed, which yields the following common structural feature: the force due to the tenth-order polynomial potential almost vanishes in the central region of each monolayer, while beyond a radial distance $\approx 63\ mm$ (the value of $R$ in (2)), the confinement force grows rapidly. As a result, particles with radial positions $\lesssim 63\ mm$ are mostly confined by an external layer of particles located at radial positions $\gtrsim 63\ mm$. As the internal structure in each case is more ordered and less dependent on the strong boundary conditions, in the following sections the diffusion and spectral analysis was performed only using the particles located within a region with radius $R_c = 50\ mm$, indicated by red circles in Fig. 2abc.

The stability of the system at each pressure is visible from the time evolution of the radial and axial velocities, as shown in Fig. 2def. For both the $5\ Pa$ and $1\ Pa$ cases, after 30s of simulation time, the horizontal velocities limit to $V_x \approx V_y \sim 10^{-5}\ m/s$, while the vertical velocities approach $V_z \sim 10^{-6}\ m/s$. The $0.1 Pa$ case exhibits distinct behavior as particle velocities experience a rapid increase to values $\sim 10^{-4}\ m/s$, with large fluctuation in both the vertical and horizontal directions, suggesting the onset of an instability. As shown in Fig. 2f the vertical velocity increases steadily for the first 14s, followed by a rapid growth which begins to saturate in the last 5 s. The growth in the horizontal velocities occurs $\approx 5s$ after the beginning of the growth in $V_z$. The onset of this instability results from the interplay between high dust mobility due to the low pressure and out-of-phase dust oscillations in the $z$-direction due to the assumed variations in dust particle size. Similar phenomenon has been observed experimentally by Gogia and Burton [56], where a dusty plasma monolayer exhibited bistability at low pressure, which was attributed to dust mass variation of similar magnitude as the variation assumed in our simulation.

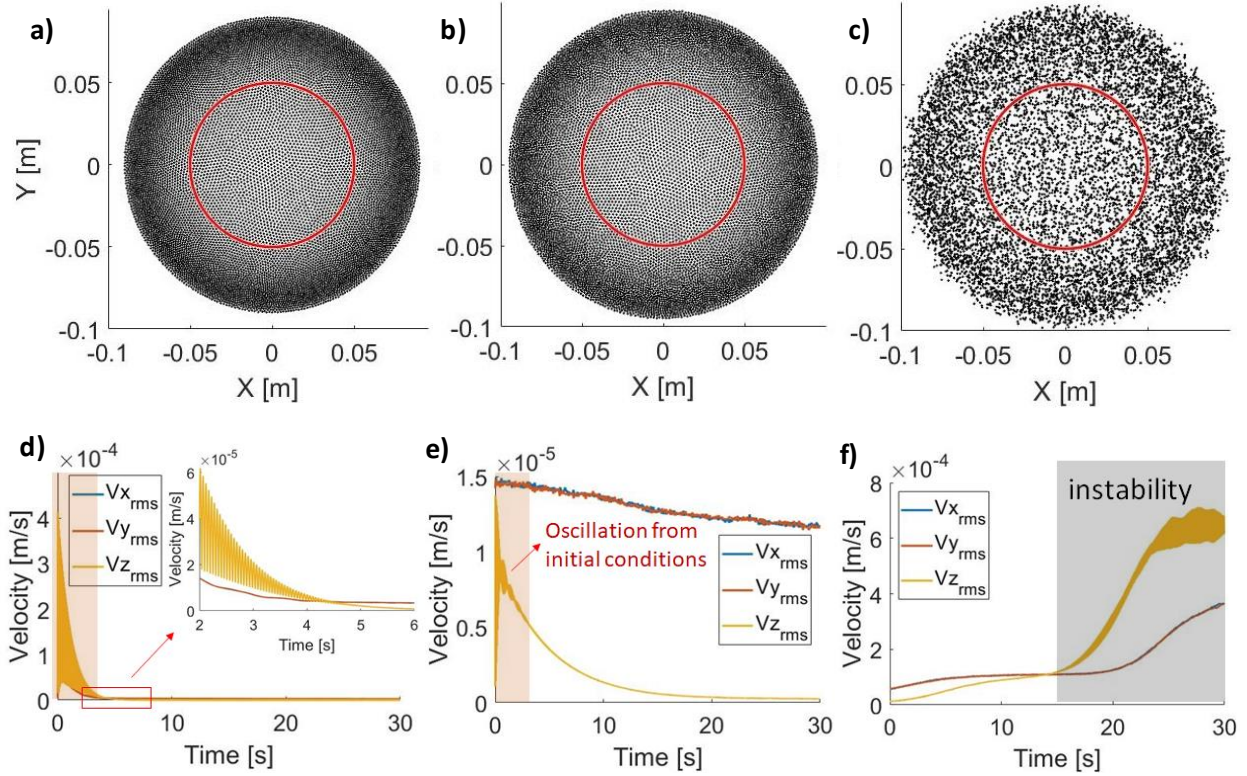

Fig. 2. Time-evolved radial positions of dust particles for a) 5 $Pa$, b) 1 $Pa$, and c) 0.1 $Pa$. The red circle in each plot indicates the central region ($R_c = 50\ mm$) of particles used for the diffusion and spectral analysis. Plots of the horizontal and vertical velocities as a function of simulation time are shown for d) 5 $Pa$, e) 1 $Pa$, and f) 0.1 $Pa$. Shaded areas indicate regions where large velocity changes are observed.

A. Crystallinity and Disorder Concentration

Dust grains in laboratory dusty plasma monolayers tend to self-organize into lattices of triangular symmetry, i.e., on average, each dust particle is located in the center of a hexagonal cell with six nearest neighbors located at each vertex. In Sec. II, we pointed out that the present simulations of dusty plasma monolayers assume particle diameters that can vary from the mean by $\pm 0.9\%$, which is typical for laboratory dusty plasma experiments. This size variation leads to variation of both dust mass and charge, which in turn, affects the spatial distribution of particles and introduces lattice defects throughout the monolayer. Two-dimensional (2D) lattice defects can be defined as the fraction of particles with number of nearest neighbors $NN$ different than six. Commonly, these are particles with five or seven nearest neighbors, located around defect lines. To quantify the amount of lattice defect in each considered monolayer, we employ a crystallinity code by Hartmann *et al.* [30], which uses a Delaunay triangulation algorithm to calculate the number of nearest neighbors for each dust grain and the complex bond-order parameter

$$G_6(i) = \frac{1}{6}\sum_{l=1}^{NN_i} exp(i6\Theta_i(l)), \qquad (3)$$

where, $NN_i$ is the number of nearest neighbors of the $i^{th}$ particle and $\Theta_i(l)$ is the angle of the $l^{th}$ nearest-neighbor bond measured with respect to the $X$-axis. When the code is applied to dust fluid structures with badly defined primitive cells, the Delaunay triangulation function returns an error due to insufficient number of unique points or overlapping particles. In other words, in these structures, the function encounters numerous points lying on the same line, in which case the triangulation does not exist. This is used to distinguish between monolayers with well-defined crystalline structure and monolayers with liquid-like structure.

Figure 3 shows the complex argument $Arg(G_6)$, which measures the overall angular orientation of a given particle neighborhood. The three plots in Fig. 3 are obtained from dust particle positions at $t = 5\ s$, at which time the $5\ Pa$ and $1\ Pa$ cases have reached equilibrium and the instability has not yet developed in the lowest pressure case (see Fig. 2). In each plot, dust particles with six nearest neighbors are marked by black dots, while dust particles with seven or more (five or less) are marked by circles (triangles). The plots of the complex argument show well-defined crystalline domains in the $5\ Pa$ and $1\ Pa$ cases, while long-distance order is lost in the $0.1Pa$ realization. In addition, the fraction of particles with $NN \neq 6$ increases as pressure decreases, yielding high defect concentration at $0.1\ Pa$.

Since the monolayer at $0.1\ Pa$ does not preserve well-defined hexagonal cells throughout the simulation, it is more appropriate to define a one-dimensional (1D) disorder of the form

$$c \approx \frac{\sigma_a}{a}, \tag{4}$$

where $\sigma_a$ is the standard deviation of the average interparticle separation $a$ computed for all dust particles within the region of interest (red circle of radius $R_c = 50\ mm$ shown in Fig. 2). We note that typical interparticle separations in this central region are $\approx 2mm$ in each pressure case, which is larger than the separation of $\approx 1.7mm$ obtained when all particles in the monolayer are accounted for. These interparticle separations are comparable to those measured for extended dusty plasma monolayers in Baylor University's Cell 3 (Fig. 1). Table II lists the 1D disorder concentration values $c$ for dust particles in the region of interest for each pressure case. These values are used as input for the Fractional Laplacian Spectral (FLS) method discussed in Sec. IV.

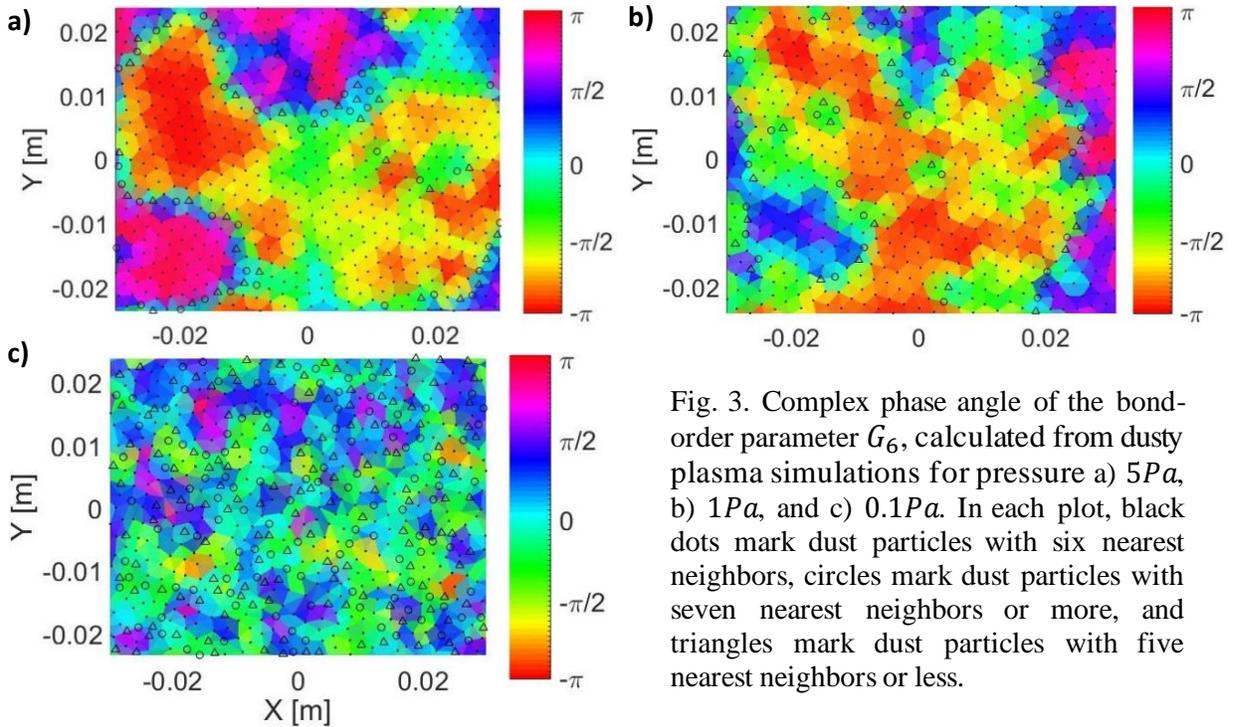

Fig. 3. Complex phase angle of the bond-order parameter $G_6$, calculated from dusty plasma simulations for pressure a) $5Pa$, b) $1Pa$, and c) $0.1Pa$. In each plot, black dots mark dust particles with six nearest neighbors, circles mark dust particles with seven nearest neighbors or more, and triangles mark dust particles with five nearest neighbors or less.

## B. Onset of Turbulence in the 0.1 $Pa$ Case

In this work, although we consider passive dust particles, we note that meaningful parallels can be drawn between the features of turbulence observed in our simulations at pressure 0.1 $Pa$ and turbulence in active matter. Active matter is commonly defined as a collection of active particles, each of which can convert the energy coming from their environment into directed motion, therefore driving the whole system far from equilibrium [67], [68]. Active organisms tend to self-organize and develop coherent collective behavior/motions. Such systems share the feature of being intrinsically out of equilibrium as energy is constantly injected at the level of each individual entity. Active systems do not have to consist of living organisms (e.g., systems of self-propelled objects, such as camphor disks studied in [44]). All of these features are characteristic of dusty plasmas, which are non-equilibrium, driven, dissipative, and exhibit self-organization and long-distance interactions. The main difference from classical self-propelled particles is that typically dust particles in dusty plasma interact with each other via electrostatic forces/scattering events and 'respond' to gradients in the electrostatic potential in the environment. We note that, beyond the parallels drawn in the present discussion, the dynamics of active systems can be directly studied in dusty plasmas, where the particles are self-propelled by "rocket force" due to material ablation [69], or dusty plasmas with Janus active particles [70]. Simulations of such active dusty plasmas are beyond the scope of the present work but represent and interesting direction for future work.

A main characteristic of 2D turbulence and turbulence in active matter is the presence of large-scale correlations. In [44], the transition from uncorrelated to interacting regime was achieved by increasing the number of active swimmers in the system to $N_p = 20 - 60$, while the system was found to freeze for $N_p \gtrsim 60$. In our simulation of dusty plasmas, the charged dust grains are interacting at $p = 0.1$ $Pa$, while the system is observed to freeze at $p \gtrsim 1$ $Pa$. The transition from an uncorrelated to a correlated regime is visible in the mean squared displacement of the particles plotted as a function of time delay, which develops a linear (diffusive) region between the ballistic region (short time scales) and the region where the MSD oscillates until it approaches a constant value (long time scales). Such dynamics are characteristic of passive particles following Lagrangian dynamics and for particle tracers in inertial turbulence, as discussed in [44]. This feature can be seen in Fig. 4a, where quadratic (purple line) and linear (red line) functions were fitted to the MSD($\tau$) plot obtained for particles in the 0.1 $Pa$ simulation. The quadratic function yields the best fit to the data in the region $\tau \in (0, 3)$ $s$, while the linear function yields the best fit in the intermediate region $\tau \in (3, 10)$ $s$. For $\tau \gtrsim 10$ $s$, the MSD plot deviates from a linear function fit and oscillates. In the present work, instead of dividing the MSD plot into individual regions and treating the dynamics in each region separately, we fit the entire curve with a power law (blue line in Fig. 4a), which allows us to treat the dynamics as anomalous diffusion, as further discussed in Sec III C. However, the presence of these regions within the MSD plot is clearly visible in Fig. 4a, which suggests transition from uncorrelated to correlated regime in the 0.1 $Pa$ case for time delays $\tau \gtrsim 10$ $s$.

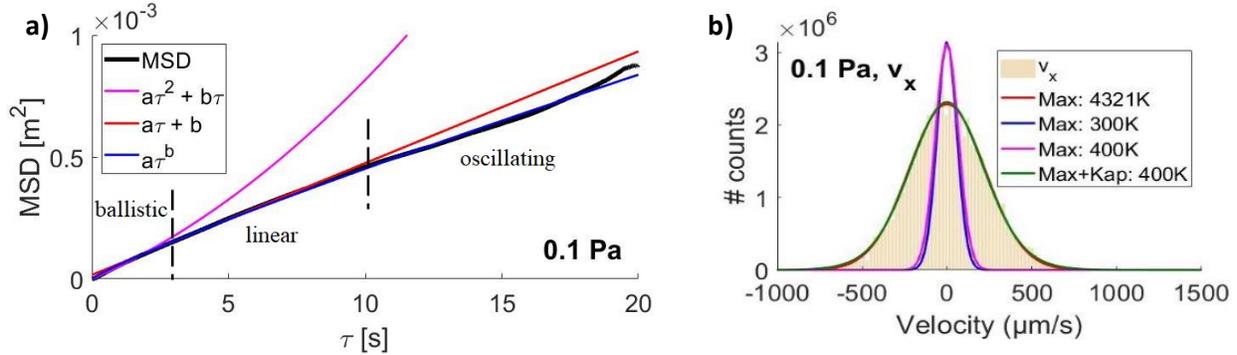

Fig. 4. a) $MSD(t)$ for a dusty plasma monolayer at pressure $0.1\ Pa$, along with quadratic, linear, and power law fits represented by the purple, red, and blue lines, respectively. b) Particle velocity distribution in the $x$ direction, fit with Maxwellian distributions (assuming different dust temperatures) and a modified Maxwellian with Kappa distribution tails (assuming a fixed temperature of $400\ K$).

Another feature, commonly reported in 2D fluid turbulence [71]–[73], is that the PDFs of velocities in the 2D plane are observed to deviate from a Gaussian. Figure 4b shows the histogram of velocities in the $x$-direction (the histogram for $v_y$ has a similar shape) for the $0.1\ Pa$ simulation, which is found to deviate from a Gaussian, and is best described by a modified Maxwellian with Kappa distribution tails. The parameter $\kappa$ characterizes how far a system is from thermal equilibrium. When $\kappa \to \infty$, a Kappa distribution approaches a Maxwellian, but when $\kappa$ is finite, the distribution function has high-energy tails, suggesting an increased number of high-velocity particles. For $\kappa < 10$, the Kappa distribution has a power law tail [74]. Here, the best fit parameter yields $\kappa \approx 11$ for the PDFs of both $v_x$ and $v_y$, suggesting that the 0.1 Pa case is not at thermal equilibrium but neither does it exhibit power-law tails. Further discussion on the velocity distributions, fits for each case, and the relation to dust temperature is included in Appendix B.

As discussed in the previous section, an instability occurs in the dusty plasma monolayer at $p = 0.1 Pa$ (Fig. 2f), which suggests the possibility of turbulent dynamics. As this instability develops at later simulation times ($t \gtrsim 14\ s$), it is not attributed to fluctuations due to initial conditions. It should be further noticed that the instability is self-excited since conditions remain unchanged during the examined time period. In the simulation, energy is imported at the individual particle level by the thermal bath and transported across larger spatial scales through dust-dust electrostatic scattering events. The increased particle mobility at low pressures allows for both enhanced dust oscillations in the vertical direction and enhanced transport in the 2D plane, which facilitates the electrostatic scattering events. When deflected from their equilibrium positions, dust particles in dusty plasmas convert electrostatic potential energy into kinetic energy, resulting in directed motion (often in the form of a restoring force). At high pressures, this kinetic energy is dissipated due to collisions with the neutral gas particles and the dust grains oscillate around their equilibrium positions with small amplitudes. At low pressures, neutral gas damping is insufficient to dissipate the energy, resulting in increased kinetic energy of the dust particles. However, due to electrostatic confinement forces and Yukawa interactions among the dust particles, the monolayer does not become dilute or gas-like. Instead, the observed state has features similar to those of 2D turbulence observed in active systems.

The above discussion implies that the most probable direction of energy flow in the simulated dusty plasma monolayers is from smaller to larger scales, i.e., an inverse energy cascade, commonly observed in active matter [43], [44]. Bourgoin *et al.* [44] showed that in the presence of turbulence in active matter, the second-order Eulerian structure function $S_E^2$ exhibits a Kolmogorovian scaling ($S_E^2 \sim r^{2/3}$) over slightly less than one decade of scales $r$. To determine if the instability in the $0.1 Pa$ case is turbulent, we computed the second-order Eulerian structure function given by

$$S_E^2(r) = \langle |(\boldsymbol{V_{ij}}(t) \cdot \boldsymbol{r_{ij}})/r_{ij}|^2 \rangle, \tag{5}$$

where $r$ is a bin of spatial scales, and the average $\langle \cdot \rangle$ is taken over all pairs $(i,j)$ of particles with separation $r_{ij}$ within that bin. The velocity difference $\boldsymbol{V_{ij}}(t) = \boldsymbol{V_i}(t) - \boldsymbol{V_j}(t)$ between each particle pair $(i,j)$ was computed using the horizontal and vertical components of the velocity vectors. As can be seen from equation (5), the structure function calculates the dissimilarity among particles velocities as a function of their spatial separation. At the onset of the instability, large velocity changes occur at small scales as the particles approach each other at distances smaller than the Debye length, which results in electrostatic scattering events. Prior to the onset of the instability, the structure function is a flat line because the particles do not approach each other at distance much smaller than the average interparticle separation. This can be seen in Fig. 5ab, which shows the structure function obtained at the 10$^{th}$ second (prior to the onset of the instability) and at the 15$^{th}$ second (after the onset of the instability).

Figure 5c shows a log-log plot of the structure function $S_E^2(r)$ obtained from position and velocity data for the $0.1 Pa$ case. The data was partitioned in bins of size $\approx 50 \mu m$, which is a few times larger than the mean dust diameter ($10 \ \mu m$), but much smaller than the Debye length ($1 \ mm$). The structure function was calculated using all 10,000 particles and averaging over 10 frames, which yields the error bars on the plot. The resulting structure function (Fig. 5c) exhibits a power law dependence $S_E^2 = ar^b$ in the interval $270 \ \mu m < r < 870 \ mm$, (shaded in the plot) suggesting direct energy cascade at these small scales. A power law fit of the form $ar^b$ was performed using the points within the region $270 \mu m - 870 \mu m$, which yields an exponent $b = 0.6628$ and a constant $a = 2.937 \times 10^6$ with $R^2 = 0.9$. Since $2/3 \approx 0.6667$, the result is reasonably close to the expected $r^{2/3}$ dependence.

At larger scales, the structure function tends to a constant asymptotic value $S_E^2 \approx 2.4 \times 10^8 \ \mu m^2/s^2$, which is expected for uncorrelated particles. This shape of the structure function is very similar to the one reported in [44] (Fig. 4a in that paper) for self-propelled interfacial particles exhibiting active turbulence. The extent of the region which exhibits $S_2 \sim r^{2/3}$ dependence suggests that the energy transfer occurs when a pair of particles is separated by less than half the interparticle separation (typically $2 \ mm$), which also coincide with the choice of Debye length $\lambda_D = 1 \ mm$ for this simulation.

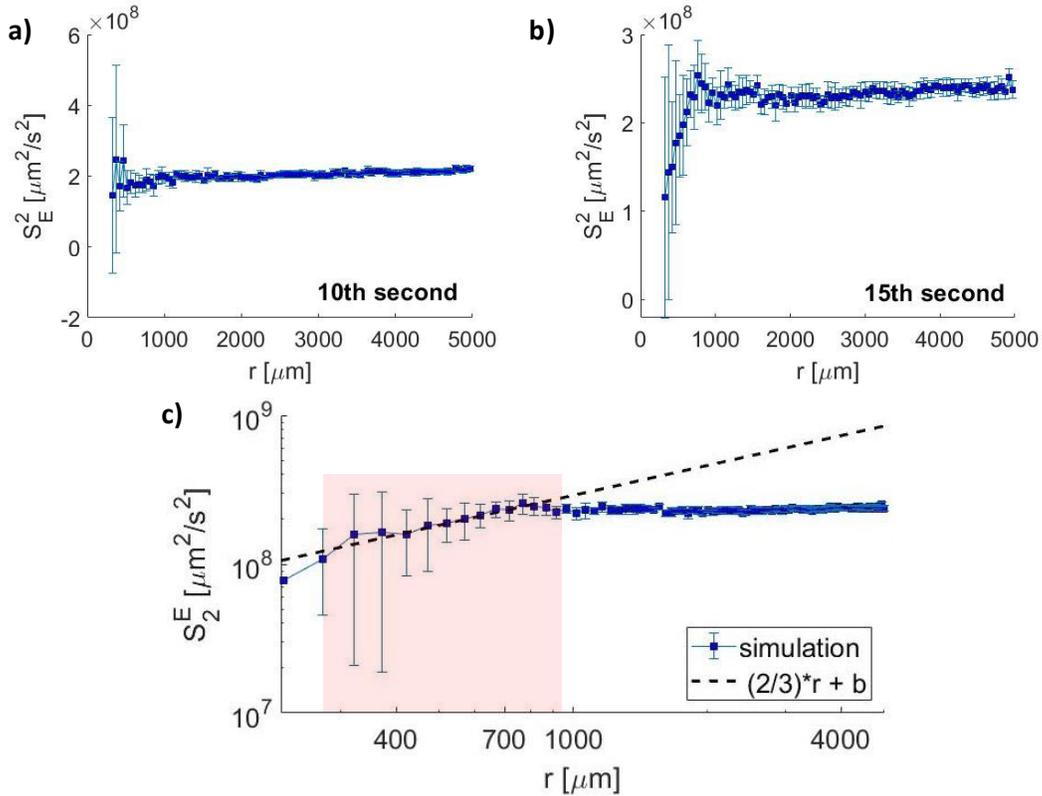

Fig.5. Second-order Eulerian structure function $S_E^2$ from particle positions and velocities obtained in the 0.1 $Pa$ simulation: a) $S_E^2$ at $t = 10\ s$, prior to the onset of the instability, b) $S_E^2$ at $t = 15\ s$, after the onset of the instability, c) log-log plot of $S_E^2$ at $t = 15\ s$. The dashed line in c) represents a fit $S_E^2 \sim r^{2/3}$ to data in the shaded region $(270 - 870)\ \mu m$. All plots were obtained using all 10,000 particles and averaging data from 10 successive frames, which yields the error bars.

From the structure function calculation, we also established that no two particles approach each other at a distance smaller than $r_{ij} \approx 300\ \mu m$. This is different from what is expected for classical self-propelled particles that can collide with each other in the absence of charge. In a typical dusty plasma experiment, each negatively charged dust grain is surrounded by a cloud of positively charged ions, which provides effective screening of the dust particle charge, with characteristic scale of this screening given by the Debye length $\lambda_D$. It is expected that for spatial scales $\gg \lambda_D$, dust grains behave as uncorrelated particles, while for scales $\lesssim \lambda_D$, correlations become significant. Therefore, it is also reasonable to observe that the energy cascade occurs at scales much bigger than the dust diameter but comparable or slightly smaller than the Debye length.

C. Dust Particle Diffusion

As discussed in the previous section, large-scale correlations determined from the Debye length in a dusty plasma can be related to the observed energy cascade driving the system's dynamics. One manifestation of correlation effects is the observation of anomalous particle diffusion, i.e., deviations from Brownian (uncorrelated) motion. A common method for assessing particle diffusion is computing the mean squared displacement (MSD) as a function of time delay $\tau$. The MSD for an individual particle is given by

$$MSD_i(\tau) = \sum_t [r_i(t+\tau) - r_i(t)]^2. \qquad (6)$$

Equation (6) can be computed and averaged over all particles in the ensemble and plotted as a function of $\tau$. In the normal diffusion regime, characteristic of uncorrelated particles, the mean square displacement (MSD) of the particle ensemble increases linearly with time, i.e., $\langle x^2 \rangle \sim t^\alpha$ where $\alpha = 1$. Deviations from normal diffusion correspond to a nontrivial topology of the corresponding phase space and spatiotemporal correlations [75]. As a result, exponents $\alpha \neq 1$ are also possible, yielding two distinct examples of anomalous transport: subdiffusion when $\alpha < 1$ and superdiffusion when $\alpha > 1$. Anomalous diffusion has been experimentally observed in various strongly coupled fluids, including ultracold neutral plasma [76], 2D and quasi-2D Yukawa liquids [77]–[80], and dusty plasmas [81]–[83]. It has been predicted that the character of the observed diffusion is sensitive to the screening length, coupling strength, energy dissipation, and choice of measurement of time scales for strongly-coupled charged liquid [84].

Another approach to identifying the presence of anomalous diffusion is to examine the probability distribution function (PDF) of the particle displacements or velocities. Normal diffusion of uncorrelated particles is characterized by a normally distributed PDF (e.g., Gaussian or Maxwellian), while anomalous diffusion is associated with non-Gaussian statistics. Using numerical techniques from Tarantino *et al.* [66], we computed the MSD plots and velocity histograms for each pressure case obtained from position and velocity data for particles within the central region of interest, as shown in Fig. 6. The number of particle trajectories detected in the central region is $\approx 2,000$ for the $5\,Pa$ and $1\,Pa$ cases and $\approx 4,000$ for the $0.1\,Pa$ case. The increased number at the lowest pressure is attributed to increased dust mobility, yielding particles leaving and entering the central region throughout the selected time period of $20\,s$, or $2,000$ frames. Thus, $\sim 10^6$ data points are considered for the statistics in each case. As shown in Figs. 6abc, the MSD in all three cases deviates from linear growth and is better approximated by a power law fit. Table II lists the extracted exponents from the best fits of the form $MSD \sim \tau^\alpha$.

Table II. Parameters measured from structural and diffusion analysis of the dusty plasma monolayers. Highlighted in red are parameters used as inputs in the FLS analysis in Sec. IV. $R_c$ is the radius of the central region of particles used in the analysis (see Fig. 2), $a$ is the average interparticle spacing, $\sigma_a$ is the standard deviation of $a$, the exponent $\alpha$ is extracted from the fit to the MSD plot, $c$ is the dimensionless disorder from equation (4), $s$ is the exponent of the fractional Laplacian operator, and $range$ is the extent of non-local interactions (Sect. IV.B.).

| Pressure [Pa]            | 5    | 1    | 0.1  |
|--------------------------|------|------|------|
| $R_c$                    | 0.05 | 0.05 | 0.05 |
| $a\,[\times 10^{-3}\,m]$ | 2.1  | 2.2  | 2.3  |
| $\sigma_a\,[\times 10^{-7}\,m]$ | 1.1 | 5.6 | 31 |
| $\alpha$                 | 1.19 | 0.83 | 0.89 |
| $c\,[\times 10^{-4}]$    | 0.5  | 2.5  | 13   |
| $s \sim 1/\alpha$        | 0.84 | 1.21 | 1.12 |
| **range** | 200 | 300 | 300 |

At the highest pressure of $5\ Pa$, the MSD fit grows slightly faster than linearly, yielding an exponent $\alpha \approx 1.19$, which suggests enhanced probability for superdiffusion. However, notice that the vertical axis scale is in units of $\sim 10^{-7}\ m^2$ and the maximum difference between the linear fit and the power law fit (at $t = 20\ s$) is on the order of $\sim 10^{-8}\ m^2$, which is two orders of magnitude smaller than the area of a typical elementary cell $\sim 10^{-6}\ m^2$ (determined by the distance to the nearest neighbors, which is $a \sim 10^{-3}\ m$). Thus, although $\alpha > 1$, superdiffusion most likely will not be observable in a laboratory experiment. This conclusion is further supported by the velocity histograms for this case (Fig. 6d), which show a minimal spread of particle velocities around the zero value.

In the $1\ Pa$ case, Fig. 6b shows that the MSD grows slower than linearly, and a power law fit to this plot yields exponent $\alpha \approx 0.83$, suggesting subdiffusive behavior. The velocity histograms (Fig. 6e) for this case show a small spread around the zero value and the maximum difference from linear dependence in the MSD plot is $\sim 10^{-7}\ m^2$ (at $t = 20\ s$, Fig 6b). As this difference is again smaller than the area of a typical elementary cell within the structure, we expect that, although at $1\ Pa$ the particles are overall more mobile, they do not exhibit vastly different dynamics than the $5\ Pa$ case.

The diffusive dynamics obtained using the $0.1\ Pa$ data are markedly different. The MSD growth (Fig. 6c) is slower than linear with a power law fit coefficient $\alpha \approx 0.89$, indicating the possibility of subdiffusion. The difference from linear growth at long times in this case is $\sim 10^{-4}\ m^2$, which exceeds the typical area of an elementary cell size by two orders of magnitude. This suggests the possibility for large displacements across the structure, which is supported by the observed tail-broadening of the velocity histograms (Fig. 6f).

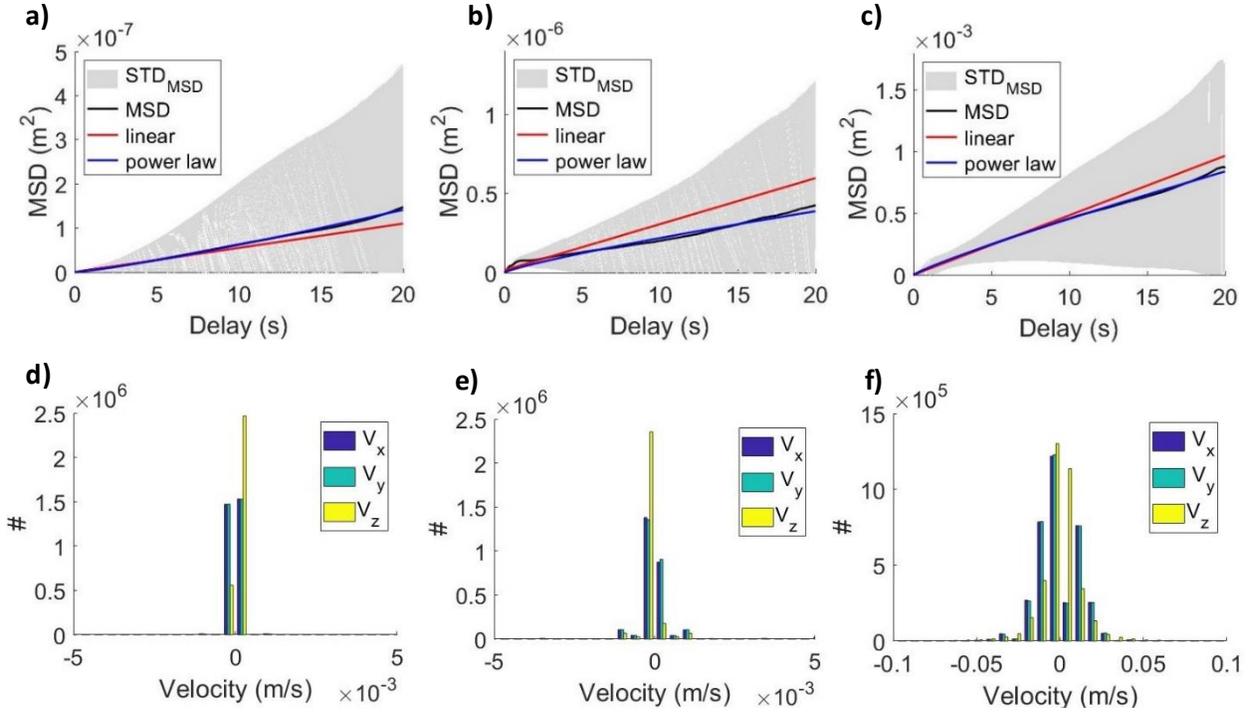

Fig. 6. Mean squared displacement (MSD) as a function of time delay for a) $5\ Pa$, b) $1\ Pa$, and c) $0.1\ Pa$ The gray shaded region shows the range of MSD values computed for all of the particles, with the average MSD shown by the black line. In each plot, a linear fit to the data is indicated by a red line, while a power law fit to the data is represented by a blue line. The corresponding velocity histograms are shown for d) $5\ Pa$, e) $1\ Pa$, and f) $0.1\ Pa$.

The conclusions drawn from the MSD plots and velocity histograms are confirmed when plotting the trajectories of all particles over the simulation time period ($30\ s$ or 3,000 frames). Figure 7ab shows the trajectories for all particles in the central region of interest over the period of $30s$ for pressures $5\ Pa$ and $1\ Pa$. As can be seen, in the $5\ Pa$ case, dust particles are arranged in a triangular lattice and the particle displacements barely deviate from their equilibrium positions. At $1\ Pa$, the dynamics are very similar to the $5\ Pa$ case, but small excitations at the individual cell level can be observed. In the $0.1\ Pa$ case, the dust particles can have large displacements and the particle trajectories completely fill in the region of interest. For clarity, Fig. 7c only shows the trajectories of 100 particles within the region of interest, which is sufficient to demonstrate the different scale of particle motion in this case. Figures 7def show the 'average' particle trajectory in each case, which is computed by averaging the displacements of all particles at each timestep. The average trajectories for $5\ Pa$ and $1\ Pa$ seem typical for a Brownian-like motion, while the trajectory computed from the $0.1\ Pa$ data resembles features of Lévy flights.

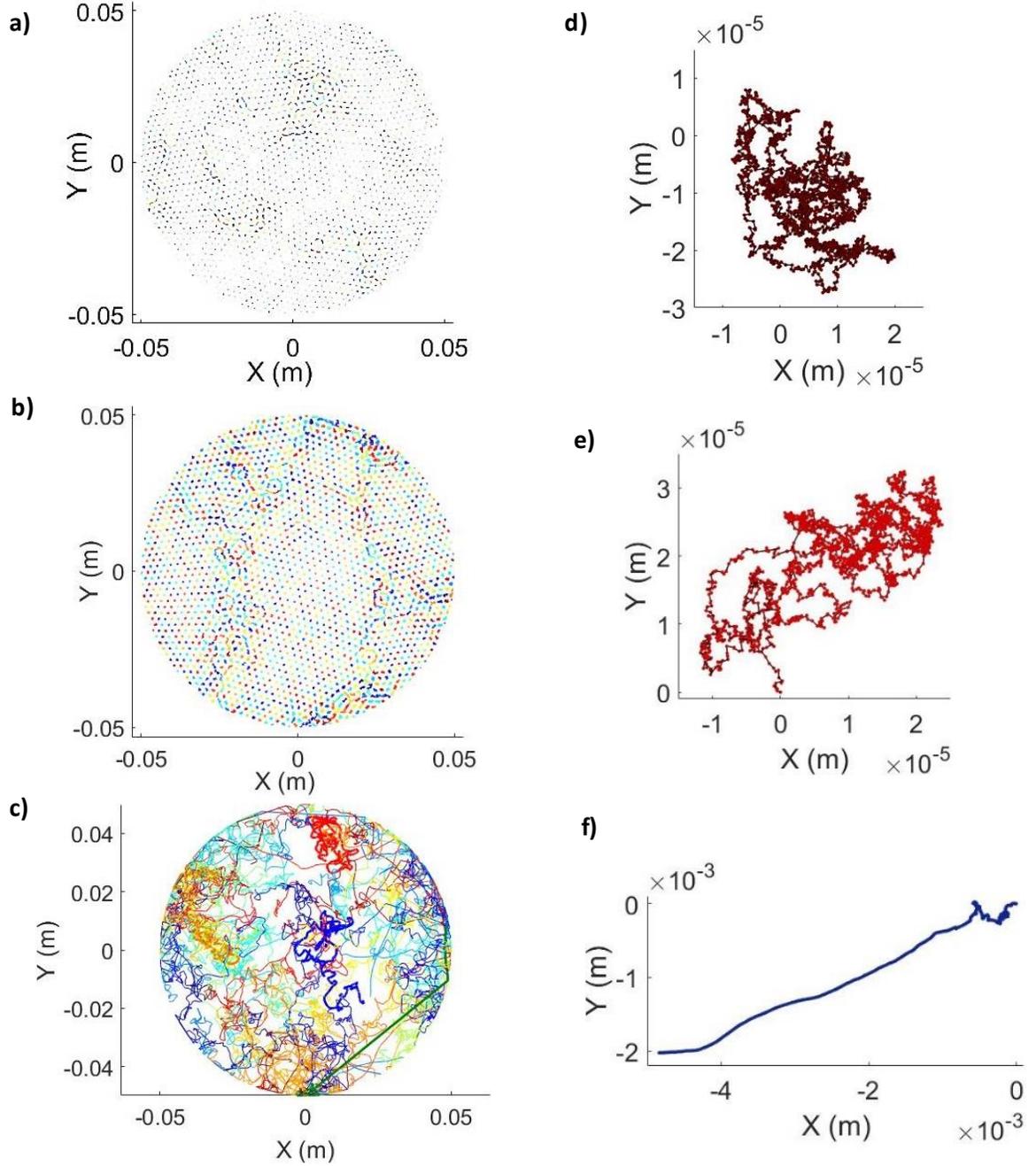

Fig. 7. Dust particle trajectories during the entire simulation time (30 $s$) for a) 5 $Pa$, b) 1 $Pa$, and c) 0.1 $Pa$. Average trajectories generated from the average displacements of all particles in each frame for d) 5 $Pa$, e) 1 $Pa$, and f) 0.1 $Pa$.

IV. ENERGY TRANSPORT

A. Fractional Laplacian Spectral (FLS) Approach

The spectral approach is a mathematical technique that determines the existence of a continuous component in the energy spectrum of an Anderson-type Hamiltonian of the form defined in [51]. The authors have previously applied the spectral approach to the study of transport in disordered 1D, 2D, and 3D lattices [59]-[62], [88], [89], where the amount of stochastic disorder $c$ was varied in the potential energy term of the examined Hamiltonian. The recently developed Fractional Laplacian Spectral (FLS) method [50], [85] applies the spectral approach to a Hamiltonian with random disorder in the potential energy term and a kinetic energy term represented by the discrete fractional Laplacian, which allows for modeling disordered media with nonlocal interactions.

In this paper we are interested in energy transport through a dusty plasma monolayer, which exhibits both disorder (discussed in Sec. III A, Eq. (4)) and anomalous diffusion due to nonlocal interactions (discussed in Sec. III C). To model energy transport in such system, we consider the *random fractional discrete Schrödinger operator*

$$H_{s,\epsilon} := (-\Delta)^s + \sum_{i \in \mathbb{Z}} \epsilon_i \langle \cdot, \delta_i \rangle \delta_i , \qquad (7)$$

where $(-\Delta)^s$, $s \in (0,2)$ is the discrete fractional Laplacian, $\delta_i$ is the $i$th standard basis vector of the 1D integer space $\mathbb{Z}$, $\langle \cdot, \cdot \rangle$ is the $\ell^2(\mathbb{Z})$ inner product, and $\epsilon_i$ are independent variables, identically distributed according to a uniform (flat) distribution on the interval $[-c/2, c/2]$, with $c > 0$. It has been shown [86], [87] that the nonlocal characteristics of anomalous diffusion can be modeled using a fractional Laplacian operator $(-\Delta)^s$, where $s \in (0,2)$. It is expected that in the subdiffusive regime ($s > 1$) transport is suppressed due to negative correlations, while for $s < 1$, transport is enhanced due to positive correlations. Padgett *et al.* [50] obtained the following 1D series representation of $(-\Delta)^s$ for the interval $s \in (0,2)$

$$(-\Delta)^s u(n) = \sum_{m \in \mathbb{Z}; m \neq n} \big(u(n) - u(m)\big) K_s(n-m) , \qquad (8)$$

where

$$K_s(m) = \begin{cases} \dfrac{4^s \Gamma(1/2+s)}{\sqrt{\pi}|-\Gamma(-s)|} \cdot \dfrac{\Gamma(|m|-s)}{\Gamma(|m|+1+s)}, & m \in \mathbb{Z} \setminus \{0\} , \\ 0, & m = 0 . \end{cases} \qquad (9)$$

Here $u$ is a discrete function on $\mathbb{Z}$ and $\Gamma$ is the standard Gamma function. The series representation of the fractional Laplacian has been recently extended to arbitrary order in [88]. Numerical simulations by Padgett *et al.* [50] provided confirmation that the representation in (8) and (9) yields enhanced transport (superdiffusion) for $s \in (0,1)$ and enhanced localization (subdiffusion) for $s \in (1,2)$, when compared to the classical case $s = 1$. The present work considers the Hamiltonian in (7) with a fractional Laplacian given by (8) and (9). The main inputs in these equations are the fractional power of the Laplacian $s$ and the disorder concentration $c$ (that determines the variables $\epsilon_i$). In Sec. III A, we obtained the value of the dimensionless disorder $c$ using equation (4). The value $s$ is known to asymptotically approach $\sim 1/\alpha$, where $\alpha$ is the exponent extracted from the MSD plots, discussed in Sec. III C and listed in Table II.

Once the appropriate Hamiltonian $H_{s,\epsilon}$ is defined, the time evolution of the initial state of the system is generated under the iterative application of $H_{s,\epsilon}$. Since this paper considers dusty plasma monolayers where energy is imported to the system at the individual particle level, we define the diameter of a single dust particle ($\approx 10 \ \mu m$) as the smallest relevant spatial scale. Larger scales are represented as multiples of this elementary scale. Physically, this means that the dust particles are viewed as rigid spheres that cannot interpenetrate and that need to move a distance of at least one dust diameter to be considered in a new spatial location. Smaller displacements are viewed as disorder, or slight deviations from the initial position. Mathematically, this amounts to viewing the standard basis vectors $\{\delta_i\}$ of the 1D integer space $\mathbb{Z}$ as discrete scales proportional to one dust diameter. We let the initial state of the system be given by $\delta_0$, located at the origin of the 1D integer space $\mathbb{Z}$, and assign to this state a normalized initial energy equal to 1. The question of interest is *how does this energy spreads across scales under the iterative application of the Hamiltonian?*

The time evolution of the initial state $\delta_0$ under the action of the Hamiltonian $H_{s,\epsilon}$ is given by the sequence $\{\delta_0, H_{s,\epsilon}\delta_0, H_{s,\epsilon}^2\delta_0, \ldots, H_{s,\epsilon}^N\delta_0\}$, where $N \in \mathbb{N}$ is the number of timesteps (equivalently, $N$ is the number of iterations of $H_{s,\epsilon}$). Let $\{\varphi_0', \varphi_1', \varphi_2', \ldots, \varphi_N'\}$ be the sequence of $\ell^2(\mathbb{Z})$ orthogonal vector states obtained from Gram-Schmidt orthogonalization of the original sequence $\{\delta_0, H_{s,\epsilon}\delta_0, H_{s,\epsilon}^2\delta_0, \ldots, H_{s,\epsilon}^N\delta_0\}$. This step allows for a proper definition of a mathematical distance between subspaces in the Hilbert space. Then, for any nontrivial vector $\nu \neq \delta_0$ (representing any spatial scale of interest), we define the distance parameter (mathematical distance) as

$$D_{s,\epsilon}^N := \sqrt{1 - \sum_{k=0}^{N}\left(\frac{\langle \nu, \varphi_k'\rangle}{\|\nu\|\|\varphi_k'\|}\right)^2}, \tag{10}$$

where $\varphi_k'$ is the $k$th term of the sequence $\{\varphi_0', \varphi_1', \varphi_2', \ldots, \varphi_N'\}$. Here, $\langle \cdot, \cdot \rangle$ is the $\ell^2(\mathbb{Z})$ inner product and $\|\cdot\|^2 = \langle \cdot, \cdot \rangle$. Equation (10) was originally derived in Liaw [89], where results from spectral theory were used to verify the following conjecture:

**Extended States Conjecture:** *For an Anderson-type Hamiltonian, if one can find a nontrivial vector $\nu$, for which the limit of the distance parameter approaches a positive value as time approaches infinity, then the spectrum of the Hamiltonian includes an absolutely continuous (ac) part, which indicates the existence of extended energy states.*

The spectrum of a Hamiltonian $H$ consists of: (i) an absolutely continuous part, corresponding to extended states and (ii) a singular part, which includes discrete eigenvalues and poorly behaved transitional states (called singular-continuous part of the spectrum). If the spectrum of $H$ coincides with the singular part (i), transport in the examined problem is localized. In the presence of non-vanishing absolutely continuous part of the spectrum, de-localization occurs in the form of extended states (by the RAGE theorem, see e.g., Sec. 1.2 of [90]). In other words, if one demonstrates with positive probability that

$$\lim_{N \to \infty} D_{s,\epsilon}^N > 0, D_{s,\epsilon}^N \in [0,1] \tag{11}$$

then the time-evolved transport behavior of the system under the action of the examined Hamiltonian includes extended energy states. In [50], it was shown that the extended states conjecture also holds for the random fractional discrete Schrödinger operator in Eq. (7).

In the following section we use the criterion (11) to determine if energy transport is likely to occur at a given spatial scale. An important distinction has to be made between energy transport and energy dissipation across scales. Here energy transport represents the existence of extended (scattering) energy states, which is characteristic of ordered systems, such as crystalline lattices. In contrast, energy dissipation across scales is a process which decreases the probability of energy transport at any particular scale. Thus, in the following discussion, we interpret $\lim_{N \to \infty} D_{s,\epsilon}^N > 0$ as high probability for energy transport and the existence of ordered structures at the examined spatial scale, while $\lim_{N \to \infty} D_{s,\epsilon}^N = 0$ as high probability for energy dissipation and the existence of disordered (dissipative) structure at the examined scale.

## B. Energy Across Scales

As discussed in the previous section, the two important inputs in the FLS method are the disorder concentration $c$ and the fractional power of the Laplacian $s$, which we determined from structural and diffusion analysis of the dusty plasma monolayers at the three simulated pressure cases. The values listed in Table II were used to define a Hamiltonian appropriate for each case. Next, we calculated the distance parameter from equation (10) for increasing values of the spatial scale, represented by the vector $v$ in equation (10). The final input needed for this calculation is the physically relevant $range$ of nonlocal interactions in the dusty plasma monolayer, which allows us to determine how many basis vectors $\delta_i$ (neighboring states) are included in the calculation at each timestep.

In a typical laboratory experiment, the negatively charged dust particles cause the formation of ion clouds, or locally correlated Debye spheres. It has been argued in [74] that the Debye length assigns a type of large-scale uncertainty in position $\sqrt{\langle \Delta r^2 \rangle} \sim \lambda_D$ and the positional variance can be expressed as

$$\sigma_r^2 = \langle \Delta r^2 \rangle = \langle r^2 \rangle - \langle r \rangle^2 \cong \frac{d+1}{2} \lambda_D^2, \quad (12)$$

where $d$ is the dimension of the space. Thus, the $range$ of nonlocal interactions can be viewed as the number of standard deviations needed to quantify the expected positional variation. For a 1D case ($d = 1$), this can be expressed as $range = n\sigma_r \cong n\lambda_D$, where $n$ is the number of standard deviations. We further normalize this value by the size a single dust diameter $d_D \approx 10\mu m$ (the smallest relevant spatial scale), which yields the dimensionless value $range^* \cong (n\lambda_D)/d_D$. For the 5 $Pa$ simulation, the assumed Debye length is $\lambda_D = 0.6mm$, yielding $range^* \approx n60$. In the 1 $Pa$ and 0.1 $Pa$ cases, $\lambda_D = 1mm$, resulting in $range^* \approx n100$.

The value $range^*$ is a numerical truncation, which determines how many neighboring states are considered when computing the action of the fractional Laplacian at each timestep. We note that the application of a fractional Laplacian $(-\Delta)^s$ results in interactions that, in principle, extend to infinity. Therefore, the application of the series representation in (8) and (9) is exact if an infinite number of neighbors are accounted for at each timestep. However, in a dusty plasma monolayer, the effective range of nonlocal interactions has a characteristic length scale given by $\lambda_D$. Thus, the numerical cutoff is not unphysical. The remaining question is whether (8) and (9) provide an accurate representation of a fractional Laplacian after the cutoff. For the three cases considered in this work, the smallest values $range^* = 60, 100$ are obtained if only one standard deviation $\sigma$ is considered. The corresponding remainder removed by the truncation is $R \sim 1/(range)^2 \sim 10^{-4}$. As this remainder is small, we expect it to yield a negligible contribution to the energy transport calculation. Based on the diffusion analysis in Sec III C (see Fig 6), although a single $\sigma_r$ may be sufficient to describe the positional spread observed for the $5Pa$ case, the development of tails in the velocity distributions at lower pressures suggests that a choice of $2\sigma$ or $3\sigma$ is more appropriate.

Figure 8 shows the time evolution of the distance parameter, calculated for $N = 3,000$ timesteps, which matches the simulation time in the many-body simulations. For each set of conditions, the resulting distance plot is an average of 10 realizations of the same numerical experiment, which minimizes fluctuations due to the random distribution of disorder values $\epsilon_i \in [-c/2, c/2\,]$ in (7). For each case, the input values $c$, $s$, and $range^*$ are listed in Table II. The reference vector in equation (10) is a linear combination of $L$ number of basis vectors in the Hilbert space, with equal weights, and has the form $v = (1/L) \sum_{j=1}^{L} \delta_j$. In this representation, a single basis vector corresponds to the minimum relevant scale (approximately equal to one dust diameter, or $10\ \mu m$). Thus, increasing the number $L$ amounts to considering larger spatial scales. Here, the examined dimensionless scales are $L = [10, 1000]$, which corresponds to spatial scales $100\ \mu m - 1\ cm$.

In Fig. 8a, corresponding to conditions obtained from the $5\ Pa$ simulation, the distance values remain close to 1 for all examined spatial scales. This corresponds to high probability for energy transport across all scales, which is typical for highly ordered crystalline lattices. At $1\ Pa$, the probability for transport is slightly decreased at scales $L \in [200, 450]$ or $2\ mm - 4.5\ mm$, which are proportional to the range of nonlocal interaction $range^* = 200$ or $2\ mm$ for this case. Nevertheless, at the final timestep considered $N = 3,000$, the distance values computed for all scales in the $1\ Pa$ case never drop to zero. Instead, the minimum value, which occurs at $L = [200, 250]$, is $D_{s,\epsilon}^{3,000} \approx 0.6$. Thus, in the $1 Pa$ case, there is strong evidence that the probability for transport is still nontrivial at all scales, which suggests a well-defined crystalline structure.

From Fig. 8c, we see that for $0.1 Pa$, the distance values visibly drop from 1 for scales $L \in [100, 500]$ or $1\ mm - 5\ mm$ and approach 0 for scales $L \in [200, 400]$ or $2\ mm - 4\ mm$ (dark blue color in Fig. 8c). For vanishing values of the distance, the spectral method cannot conclude the existence of transport at the considered vector scale $v$. Thus, our interpretation of these calculations is that there is low probability for transport at the corresponding spatial scales. Instead, closer examination of the distance plots corresponding to $L \in [10, 450]$ (Fig. 8d) suggests that in this interval, the probability for transport rapidly decreases as scales increase, which can be interpreted as energy dissipation across scales. This suggests the presence of an instability in the corresponding dusty plasma monolayer, which is consistent with the observations from the many-body simulations.

## V. COMPARISON OF KINETIC AND SPECTRAL APPROACH

Since in Fig. 8 the examined dimensionless reference scales $L \in [10, 1000]$ were normalized by the dust particle diameter $d_D \approx 10\ \mu m$, the corresponding spatial scales vary in the range $100\ \mu m$ to $1\ cm$. Comparison to the cell orientation maps of Fig. 3a shows that for the $5\ Pa$ case, scales in the range $100\ \mu m - 1\ cm$ are smaller than the typical size of the observed crystalline domains (with characteristic 1D scales $\gtrsim 2 cm$). This explains the observed high probability for transport at all scales. Similarly, examination of Fig. 3b suggests that at $1\ Pa$, small sub-structures are formed within the larger crystalline domains. Those consist of individual cells with different spatial orientation or small clusters of 'defect' cells that have five or seven nearest neighbors. The 1D scale of an individual cell is approximately equal to twice the mean interparticle separation, or $\approx 2 \times 2\ mm$ for the central regions of all the examined monolayers. In Fig. 8b, the FLS method predicts decreased probability for transport at dimensionless scales $L \in [200, 450]$, corresponding to spatial range of $2\ mm - 4.5\ mm$, which is in agreement with the observation of substructure formation within the larger crystalline domains in the $1\ Pa$ simulations.

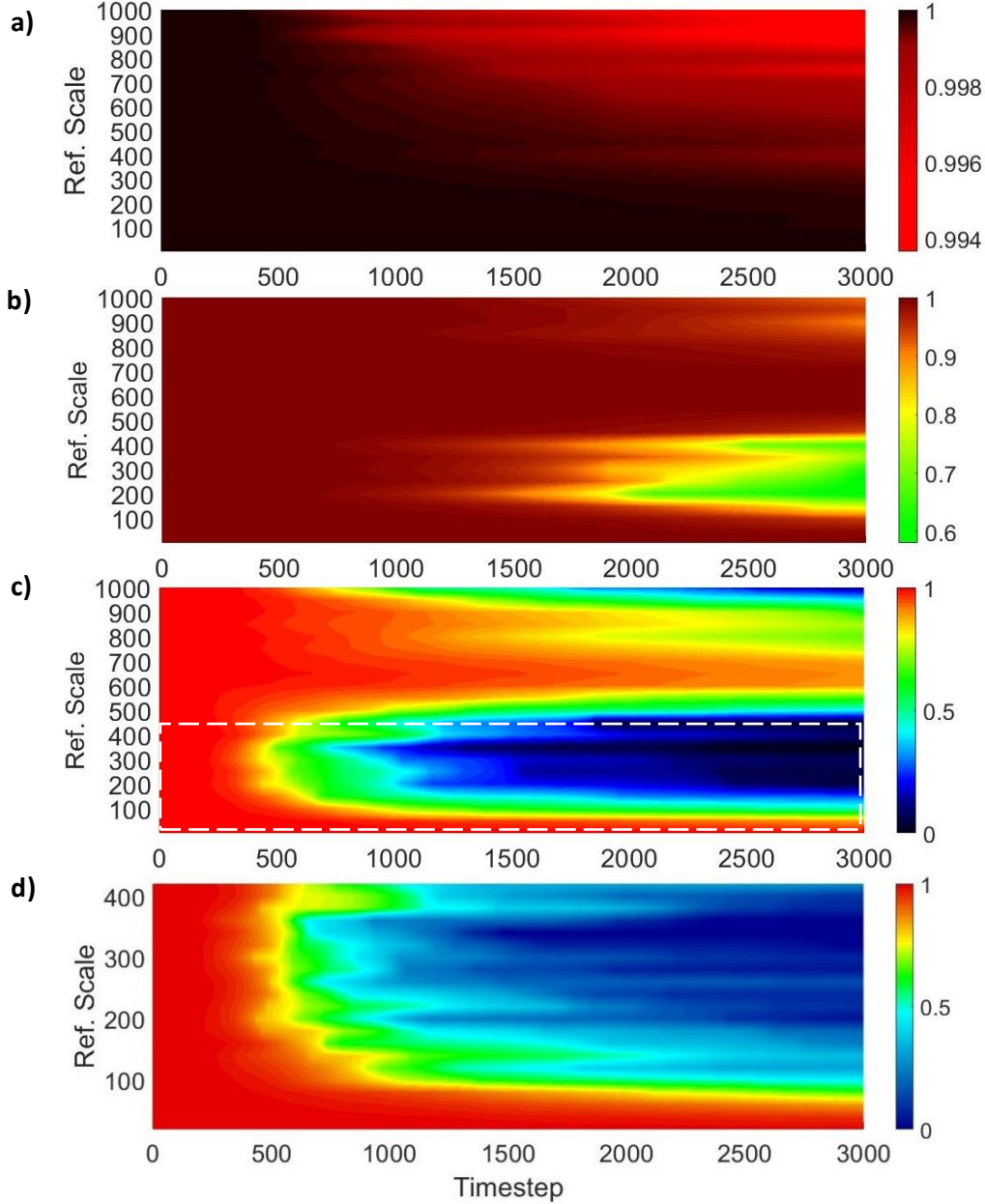

Fig. 8. Time evolution of the distance parameter from equation (10), calculated for increasing reference vector scales for the three pressure cases a) $5Pa$, b) $1Pa$, and c) $0.1Pa$. The range of nonlocal interactions in a) is 200, while the range in b) and c) is 300. Other parameters used for each case can be found in Table II. All distance calculations are averages of 10 realizations. Plots in d) correspond to the region of scales in c) marked by the white dashed rectangle.

For the $0.1\ Pa$ case, Fig. 3c clearly shows the formation of sub-cell structure as indicated by the variation of colors across portions of individual cells. Additionally, we note that this cell orientation map could only be generated in the initial timesteps of the simulation before the onset of the instability. At later timesteps, the dust particle monolayer loses crystallinity and dust positions are observed to substantially vary from the average interparticle separation. This is evident both from Fig. 7 and from the high value of the corresponding dimensionless disorder $c = 1.3 \times 10^{-3}$ listed in Table II. Fig. 8cd suggest that in the interval $L \in [200, 400]$, corresponding to spatial scales of $2\ mm - 4\ mm$, transport is *not* likely to occurs. As these spatial scales coincide with $1 - 2$ interparticle distances, this result confirms the breaking of dust ordering into well-defined cells, observed in the simulation. The higher probability for energy transport at larger scales $L \in [500, 1000]$, or $5\ mm - 10\ mm$, can be explained by the presence of global liquid-like behavior of the monolayer. In other words, although dust ordering in individual cells is lost, the structure is not chaotic and still preserves the characteristics of a strongly coupled monolayer.

## VI. CONCLUSION

In this article, we presented a theoretical study of energy transport in dusty plasma monolayers. Specifically, we considered three sets of experimentally relevant conditions, which are expected to result in a highly-ordered crystalline structure at $5\ Pa$, a disordered crystalline structure at $1\ Pa$, and a liquid-like structure at $0.1\ Pa$. The latter case is observed to develop a self-excited instability, having the characteristics of active turbulence, where energy cascades from smaller to larger scales. We modeled these three structures using a many-body simulation of 10,000 spherical dust particles, whose dust diameter was allowed to vary by $\pm 0.9\%$. The variation in particle size introduces variation of dust particle mass and charge, which affects the spatial distribution of particles within the monolayer. To mimic the confinement forces expected in large-electrode dusty plasma cells, we assumed a tenth-order radial confinement potential, which has a flat profile in the center of the simulation box and steep slope close to the exterior. The resulting dusty plasma structures form central regions of particles, where radial confinement is provided indirectly through an external later of particles. In this manner, focusing the analysis on these central regions of dust particles allows to minimize the effect of boundary conditions.

For each examined pressure case, we performed structural and diffusion analysis of the simulated dusty plasmas, which confirmed that for $5\ Pa$ and $1\ Pa$, the obtained monolayers are crystalline, while the $0.1\ Pa$ condition yields a liquid-like structure that develops a self-excited turbulent instability. For each case, we computed a dimensionless disorder $c$ from the standard deviation from the mean interparticle separation, which was found to increase inversely with pressure. The dust particle diffusion regime was identified from analysis of the mean squared displacement (MSD) as a function of time and from velocity histograms. While the particles in all three cases are found to move according to anomalous diffusion, only at $0.1\ Pa$ does the observed deviation from classical diffusion result in significant change of the individual particle trajectories. This is also confirmed by the velocity histograms, which exhibit tail broadening at this pressure. For each case, a power law fit to the plot of $MSD(\tau)$ was used to extract an exponent $\alpha$, which quantifies the deviation from classical diffusive behavior. The observation of anomalous diffusion, or $\alpha \neq 1$, suggests the presence of nonlocal interactions, which is expected for dusty particles interacting through a shielded Coulomb (Yukawa) potential. The characteristic range of such nonlocal interactions was determined from the Debye length $\lambda_D$ selected for each simulation.

To study energy transport across scales in each monolayer, we used a Fractional Laplacian Spectral (FLS) method, where the time evolution of the initial energy state of the system is obtained from the iterative application of the corresponding Hamiltonian. To account for both random disorder and nonlocal interactions, we employed a random fractional discrete Schrödinger operator as the Hamiltonian. The fractional power $s$ of the Laplacian is known to asymptotically approach $1/\alpha$, where $\alpha$ is the exponent characterizing time evolution of the MSD plots. This fraction and the range of nonlocal interactions determined from the many-body simulations were used to obtain the fractional Laplacian for each case. The dimensionless disorder calculated for each monolayer was used to define the potential term in the corresponding Hamiltonian. Using the FLS method, we computed the distance parameter in equation (10), which quantifies the probability for transport (in the form of extended energy states) as a function of spatial scale. The results showed that at $5\ Pa$, energy transport is likely to occur at all scales, in agreement with the fact that the corresponding structure is highly ordered with large crystalline domains. At $1\ Pa$, the probability for energy transport is slightly decreased at energy scales proportional to the size of individual elementary cells, which coincides with the observed formation of substructures within the large crystalline domains in that case. Finally, at $0.1\ Pa$, the FLS calculations demonstrated that energy transport is suppressed for spatial scales proportional to the size of individual cells but preserved at larger spatial scales. These results were found to coincide with the observed breaking of elementary cells as the monolayer transitioned to unstable liquid-like structure.

The possible relationship between classical inertial turbulence and 2D turbulence has been studied in monolayers of active particles [44], but still presents a relatively unexplored question in plasma physics. Here we showed that, for a dusty plasma monolayer at $0.1\ Pa$, the time evolution of the MSD plots exhibits a transition from uncorrelated to correlated regime at intermediate time scales. It was further demonstrated that the velocity histograms deviate from Maxwellian distributions by developing 'fat' tails, which suggests the anomalous diffusion of the dust particles. While both these features are expected for 2D turbulence, the presence of $r^{2/3}$ dependence in the corresponding structure function plot is a feature typical of 3D inertial turbulence. The coexistence of these features at small scales is possible due to the quasi-2D nature of dusty plasma monolayers, where random fluctuations in the $z$-displacements of the particles can result in the onset of a global instability in the monolayer plane. Additionally, electrostatic scattering among charged particles in a dusty plasma introduce a mechanism for energy transfer at scales on the order of the Debye length (much bigger than scales proportional to the particle size, where energy is transferred through collisions). Thus, dusty plasma monolayers present an appropriate platform for studying the fundamental physical mechanisms driving both direct and inverse energy cascades in complex systems, where both random disorders and nonlocal interactions may be present.

## ACKNOWLEDGMENTS

This work was supported by the NSF grant numbers 1903450 (EGK, JLP, CDL, and LSM), 1707215 (LSM and TWH), and 1740203 (TWH and LSM), NSF-DMS grant number 1802682 (CDL), and NASA grant number 1571701 (TWH and LSM), and DOE grant number SC0021284 (EGK).

Since August 2020, CDL has been serving as a Program Director in the Division of Mathematical Sciences at the National Science Foundation (NSF), USA, and as a component of this position, she received support from NSF for research, which included work on this paper. Any opinions, findings, and conclusions or recommendations expressed in this material are those of the authors and do not necessarily reflect the views of the National Science Foundation.

## DATA AVAILABILITY

The data that support the findings of this study are available from the corresponding author upon reasonable request.

**Bibliography**

[1] J. Lu and R. A. Shaw, "Charged particle dynamics in turbulence: Theory and direct numerical simulations," *Phys. Fluids*, vol. 27, no. 6, p. 065111, Jun. 2015, doi: 10.1063/1.4922645.

[2] J. K. Eaton and J. R. Fessler, "Preferential concentration of particles by turbulence," *Int. J. Multiph. Flow*, vol. 20, pp. 169–209, Aug. 1994, doi: 10.1016/0301-9322(94)90072-8.

[3] J. R. Fessler, J. D. Kulick, and J. K. Eaton, "Preferential concentration of heavy particles in a turbulent channel flow," *Phys. Fluids*, vol. 6, no. 11, pp. 3742–3749, Nov. 1994, doi: 10.1063/1.868445.

[4] G. E. Thomas and J. Olivero, "Noctilucent clouds as possible indicators of global change in the mesosphere," *Adv. Space Res.*, vol. 28, no. 7, pp. 937–946, Jan. 2001, doi: 10.1016/S0273-1177(01)80021-1.

[5] K. V. Beard, H. T. Ochs, and C. H. Twohy, "Aircraft measurements of high average charges on cloud drops in layer clouds," *Geophys. Res. Lett.*, vol. 31, no. 14, Jul. 2004, doi: 10.1029/2004GL020465.

[6] L. Zhou and B. A. Tinsley, "Production of space charge at the boundaries of layer clouds," *J. Geophys. Res. Atmospheres*, vol. 112, no. D11, Jun. 2007, doi: 10.1029/2006JD007998.

[7] G. Hendrickson, "Electrostatics and gas phase fluidized bed polymerization reactor wall sheeting," *Chem. Eng. Sci.*, vol. 61, no. 4, pp. 1041–1064, Feb. 2006, doi: 10.1016/j.ces.2005.07.029.

[8] R. G. Rokkam, A. Sowinski, R. O. Fox, P. Mehrani, and M. E. Muhle, "Computational and experimental study of electrostatics in gas–solid polymerization fluidized beds," *Chem. Eng. Sci.*, vol. 92, pp. 146–156, Apr. 2013, doi: 10.1016/j.ces.2013.01.023.

[9] R. G. Rokkam, R. O. Fox, and M. E. Muhle, "Computational fluid dynamics and electrostatic modeling of polymerization fluidized-bed reactors," *Powder Technol.*, vol. 203, no. 2, pp. 109–124, Nov. 2010, doi: 10.1016/j.powtec.2010.04.002.

[10] J. S. Shrimpton and A. J. Yule. Characterisation of charged hydrocarbon sprays for application in combustion systems. Experiments in fluids, 26(5), 460-469, 1999.

[11] F. Esposito, R. Molinaro, C. I. Popa, C.E.S.A.R.E Molfese, F. Cozzolino, L. Marty, K. Taj-Eddine, G. Di Achille, G. Franzese, S. Silvestro, and G. G. Ori, "The role of the atmospheric electric field in the dust-lifting process," *Geophys. Res. Lett.*, vol. 43, no. 10, pp. 5501–5508, May 2016, doi: 10.1002/2016GL068463.

[12] J. F. Kok and N. O. Renno, "The effects of electric forces on dust lifting: Preliminary studies with a numerical model," *J. Phys. Conf. Ser.*, vol. 142, no. 1, p. 012047, 2008, doi: 10.1088/1742-6596/142/1/012047.

[13] A. V. Malm and T. A. Waigh, "Elastic turbulence in entangled semi-dilute DNA solutions measured with optical coherence tomography velocimetry," *Sci. Rep.*, vol. 7, no. 1, p. 1186, Apr. 2017, doi: 10.1038/s41598-017-01303-4.

[14] J. Mitchell, K. Lyons, A. M. Howe, and A. Clarke, "Viscoelastic polymer flows and elastic turbulence in three-dimensional porous structures," *Soft Matter*, vol. 12, no. 2, pp. 460–468, Dec. 2015, doi: 10.1039/C5SM01749A.

[15] A. Groisman and V. Steinberg, "Elastic turbulence in a polymer solution flow," *Nature*, vol. 405, no. 6782, pp. 53–55, May 2000, doi: 10.1038/35011019.

[16] A. Groisman and V. Steinberg, "Elastic turbulence in curvilinear flows of polymer solutions," *New J. Phys.*, vol. 6, no. 1, p. 29, 2004, doi: 10.1088/1367-2630/6/1/029.

[17] S. Bu, J. Yang, Q. Dong, and Q. Wang, "Experimental study of flow transitions in structured packed beds of spheres with electrochemical technique," *Exp. Therm. Fluid Sci.*, vol. 60, pp. 106–114, Jan. 2015, doi: 10.1016/j.expthermflusci.2014.09.001.

[18] N. A. Horton and D. Pokrajac, "Onset of turbulence in a regular porous medium: An experimental study," *Phys. Fluids*, vol. 21, no. 4, p. 045104, Apr. 2009, doi: 10.1063/1.3091944.

[19] J. W. Fox, "Onset of Turbulent Flow in certain Arrays of Particles," *Proc. Phys. Soc. Sect. B*, vol. 62, no. 12, p. 829, 1949, doi: 10.1088/0370-1301/62/12/309.

[20] J. C. M. Lin and L. L. Pauley, "Low-Reynolds-number separation on an airfoil," *AIAA J.*, vol. 34, no. 8, pp. 1570–1577, 1996, doi: 10.2514/3.13273.

[21] M. S. Selig and J. J. Guglielmo, "High-Lift Low Reynolds Number Airfoil Design," *J. Aircr.*, vol. 34, no. 1, pp. 72–79, 1997, doi: 10.2514/2.2137.

[22] W. Shyy, Y. Lian, J. Tang, D. Viieru, and H. Liu, *Aerodynamics of Low Reynolds Number Flyers*. Cambridge University Press, 2007.

[23] Z. Jane Wang, "Two Dimensional Mechanism for Insect Hovering," *Phys. Rev. Lett.*, vol. 85, no. 10, pp. 2216–2219, Sep. 2000, doi: 10.1103/PhysRevLett.85.2216.

[24] C. P. Ellington, "The novel aerodynamics of insect flight: applications to micro-air vehicles," *J. Exp. Biol.*, vol. 202, no. 23, pp. 3439–3448, Dec. 1999.

[25] R. Golestanian and A. Ajdari, "Stochastic low Reynolds number swimmers," *J. Phys. Condens. Matter*, vol. 21, no. 20, p. 204104, 2009, doi: 10.1088/0953-8984/21/20/204104.

[26] A. Najafi and R. Golestanian, "Coherent hydrodynamic coupling for stochastic swimmers," *EPL Europhys. Lett.*, vol. 90, no. 6, p. 68003, 2010, doi: 10.1209/0295-5075/90/68003.

[27] J. H. Chu and L. I, "Direct observation of Coulomb crystals and liquids in strongly coupled rf dusty plasmas," *Phys. Rev. Lett.*, vol. 72, no. 25, pp. 4009–4012, Jun. 1994, doi: 10.1103/PhysRevLett.72.4009.

[28] H. Thomas, G. E. Morfill, V. Demmel, J. Goree, B. Feuerbacher, and D. Möhlmann, "Plasma Crystal: Coulomb Crystallization in a Dusty Plasma," *Phys. Rev. Lett.*, vol. 73, no. 5, pp. 652–655, Aug. 1994, doi: 10.1103/PhysRevLett.73.652.

[29] A. P. Nefedov, O. F. Petrov, V. I. Molotkov, and V. E. Fortov, "Formation of liquidlike and crystalline structures in dusty plasmas," *J. Exp. Theor. Phys. Lett.*, vol. 72, no. 4, pp. 218–226, Aug. 2000, doi: 10.1134/1.1320134.

[30] P. Hartmann, A. Douglass, J. C. Reyes, L. S. Matthews, T. W. Hyde, A. Kovács, and Z. Donkó, "Crystallization Dynamics of a Single Layer Complex Plasma," *Phys. Rev. Lett.*, vol. 105, p. 115004, 2010, doi: 10.1103/PhysRevLett.105.115004.

[31] M. Schwabe, S. Zhdanov, C. Räth, D. B. Graves, H. M. Thomas, and G. E. Morfill, "Collective Effects in Vortex Movements in Complex Plasmas," *Phys. Rev. Lett.*, vol. 112, no. 11, p. 115002, Mar. 2014, doi: 10.1103/PhysRevLett.112.115002.

[32] G. E. Morfill, M. Rubin-Zuzic, H. Rothermel, A. V. Ivlev, B. A. Klumov, H. M. Thomas, U. Konopka, and V. Steinberg, V., "Highly Resolved Fluid Flows: ``Liquid Plasmas'' at the Kinetic Level," *Phys. Rev. Lett.*, vol. 92, no. 17, p. 175004, Apr. 2004, doi: 10.1103/PhysRevLett.92.175004.

[33] M. Rubin-Zuzic, H. M. Thomas, S. K. Zhdanov, and G. E. Morfill, "Circulation' dynamo in complex plasma," *New J. Phys.*, vol. 9, no. 2, p. 39, 2007, doi: 10.1088/1367-2630/9/2/039.

[34] M. Klindworth, A. Melzer, A. Piel, and V. A. Schweigert, "Laser-excited intershell rotation of finite Coulomb clusters in a dusty plasma," *Phys. Rev. B*, vol. 61, no. 12, pp. 8404–8410, Mar. 2000, doi: 10.1103/PhysRevB.61.8404.

[35] G. Uchida, S. Iizuka, T. Kamimura, and N. Sato, "Generation of two-dimensional dust vortex flows in a direct current discharge plasma," *Phys. Plasmas*, vol. 16, no. 5, p. 053707, May 2009, doi: 10.1063/1.3139252.

[36] V. Nosenko and J. Goree, "Shear Flows and Shear Viscosity in a Two-Dimensional Yukawa System (Dusty Plasma)," *Phys. Rev. Lett.*, vol. 93, no. 15, p. 155004, Oct. 2004, doi: 10.1103/PhysRevLett.93.155004.

[37] R. Heidemann, S. Zhdanov, K. R. Sütterlin, H. M. Thomas, and G. E. Morfill, "Shear flow instability at the interface among two streams of a highly dissipative complex plasma," *EPL Europhys. Lett.*, vol. 96, no. 1, p. 15001, 2011, doi: 10.1209/0295-5075/96/15001.

[38] A. Gupta, R. Ganesh, and A. Joy, "Kolmogorov flow in two dimensional strongly coupled dusty plasma," *Phys. Plasmas*, vol. 21, no. 7, p. 073707, Jul. 2014, doi: 10.1063/1.4890488.

[39] S. Zhdanov, M. Schwabe, C. Räth, H. M. Thomas, and G. E. Morfill, "Wave turbulence observed in an auto-oscillating complex (dusty) plasma," *EPL Europhys. Lett.*, vol. 110, no. 3, p. 35001, 2015, doi: 10.1209/0295-5075/110/35001.

[40] Y.-Y. Tsai, M.-C. Chang, and L. I, "Observation of multifractal intermittent dust-acoustic-wave turbulence," *Phys. Rev. E*, vol. 86, no. 4, p. 045402, Oct. 2012, doi: 10.1103/PhysRevE.86.045402.

[41] J. Pramanik, B. M. Veeresha, G. Prasad, A. Sen, and P. K. Kaw, "Experimental observation of dust-acoustic wave turbulence," *Phys. Lett. A*, vol. 312, no. 1, pp. 84–90, Jun. 2003, doi: 10.1016/S0375-9601(03)00614-5.

[42] "The local structure of turbulence in incompressible viscous fluid for very large Reynolds numbers," *Proc. R. Soc. Lond. Ser. Math. Phys. Sci.*, Jul. 1991, doi: 10.1098/rspa.1991.0075.

[43] H. H. Wensink, J. Dunkel, S. Heidenreich, K. Drescher, R. E. Goldstein, H. Löwen, and J. M. Yeomans, "Meso-scale turbulence in living fluids," *Proc. Natl. Acad. Sci.*, vol. 109, no. 36, pp. 14308–14313, Sep. 2012, doi: 10.1073/pnas.1202032109.

[44] M. Bourgoin, R. Kervil, C. Cottin-Bizonne, F. Raynal, R. Volk, and C. Ybert, "Kolmogorovian Active Turbulence of a Sparse Assembly of Interacting Marangoni Surfers," *Phys. Rev. X*, vol. 10, no. 2, p. 021065, Jun. 2020, doi: 10.1103/PhysRevX.10.021065.

[45] J. K. Meyer, I. Laut, S. K. Zhdanov, V. Nosenko, and H. M. Thomas, "Coupling of Noncrossing Wave Modes in a Two-Dimensional Plasma Crystal," *Phys. Rev. Lett.*, vol. 119, no. 25, p. 255001, Dec. 2017, doi: 10.1103/PhysRevLett.119.255001.

[46] C. M. Ticoş, D. Ticoş, and J. D. Williams, "Kinetic effects in a plasma crystal induced by an external electron beam," *Phys. Plasmas*, vol. 26, no. 4, p. 043702, Apr. 2019, doi: 10.1063/1.5092749.

[47] V. Nosenko, M. Pustylnik, M. Rubin-Zuzic, A. M. Lipaev, A. V. Zobnin, A. D. Usachev, H. M. Thomas, M. H. Thoma, V. E. Fortov, O. Kononenko, and A. Ovchinin, "Shear flow in a three-dimensional complex plasma in microgravity conditions," *Phys. Rev. Res.*, vol. 2, no. 3, p. 033404, Sep. 2020, doi: 10.1103/PhysRevResearch.2.033404.

[48] C. Liaw, "Approach to the Extended States Conjecture," *J. Stat. Phys.*, vol. 153, no. 6, pp. 1022–1038, Dec. 2013, doi: 10.1007/s10955-013-0879-5.

[49] W. King, R. C. Kirby, and C. Liaw, "Delocalization for the 3-D discrete random Schroedinger operator at weak disorder," *J. Phys. Math. Theor.*, vol. 47, no. 30, p. 305202, Aug. 2014, doi: 10.1088/1751-8113/47/30/305202.

[50] J. L. Padgett, E. G. Kostadinova, C. D. Liaw, K. Busse, L. S. Matthews, and T. W. Hyde, "Anomalous diffusion in one-dimensional disordered systems: a discrete fractional Laplacian method," *J. Phys. Math. Theor.*, vol. 53, no. 13, p. 135205, Mar. 2020, doi: 10.1088/1751-8121/ab7499.

[51] V. Jakšić and Y. Last, "Simplicity of singular spectrum in Anderson-type Hamiltonians," *Duke Math. J.*, vol. 133, no. 1, pp. 185–204, May 2006, doi: 10.1215/S0012-7094-06-13316-1.

[52] J. K. Olthoff and K. E. Greenberg, "The Gaseous Electronics Conference RF Reference Cell—An Introduction," *J. Res. Natl. Inst. Stand. Technol.*, vol. 100, no. 4, pp. 327–339, 1995, doi: 10.6028/jres.100.025.

[53] V. Land, E. Shen, B. Smith, L. Matthews, and T. Hyde, "Experimental and computational characterization of a modified GEC cell for dusty plasma experiments," *New J. Phys.*, vol. 11, no. 6, p. 063024, Jun. 2009, doi: 10.1088/1367-2630/11/6/063024.

[54] H. M. Anderson and S. B. Radovanov, "Dusty Plasma Studies in the Gaseous Electronics Conference Reference Cell," *J. Res. Natl. Inst. Stand. Technol.*, vol. 100, no. 4, pp. 449–462, 1995, doi: 10.6028/jres.100.034.

[55] J. B. Pieper, J. Goree, and R. A. Quinn, "Experimental studies of two-dimensional and three-dimensional structure in a crystallized dusty plasma," *J. Vac. Sci. Technol. A*, vol. 14, no. 2, pp. 519–524, Mar. 1996, doi: 10.1116/1.580118.

[56] G. Gogia and J. C. Burton, "Emergent Bistability and Switching in a Nonequilibrium Crystal," *Phys. Rev. Lett.*, vol. 119, no. 17, p. 178004, Oct. 2017, doi: 10.1103/PhysRevLett.119.178004.

[57] L. S. Matthews, D. L. Sanford, E. G. Kostadinova, K. S. Ashrafi, E. Guay, and T. W. Hyde, "Dust charging in dynamic ion wakes," *Phys. Plasmas*, vol. 27, no. 2, p. 023703, Feb. 2020, doi: 10.1063/1.5124246.

[58] L. S. Matthews and T. W. Hyde, "Effect of dipole–dipole charge interactions on dust coagulation," *New J. Phys.*, vol. 11, no. 6, p. 063030, 2009, doi: 10.1088/1367-2630/11/6/063030.

[59] L. S. Matthews and T. W. Hyde, "Effects of the charge-dipole interaction on the coagulation of fractal aggregates," *IEEE Trans. Plasma Sci.*, vol. 32, no. 2, pp. 586–593, Apr. 2004, doi: 10.1109/TPS.2004.826107.

[60] L. S. Matthews and T. W. Hyde, "Charged grains in Saturn's F-Ring: interaction with Saturn's magnetic field," *Adv. Space Res.*, vol. 33, no. 12, pp. 2292–2297, 2004, doi: 10.1016/S0273-1177(03)00462-9.

[61] L. S. Matthews and T. W. Hyde, "Charging and Growth of Fractal Dust Grains," *IEEE Trans. Plasma Sci.*, vol. 36, no. 1, pp. 310–314, Feb. 2008, doi: 10.1109/TPS.2007.913923.

[62] L. S. Matthews and T. W. Hyde, "Gravitoelectrodynamics in Saturn's F ring: encounters with Prometheus and Pandora," *J. Phys. Math. Gen.*, vol. 36, no. 22, p. 6207, 2003, doi: 10.1088/0305-4470/36/22/349.

[63] M. Sun, L. S. Matthews, and T. W. Hyde, "Effect of multi-sized dust distribution on local plasma sheath potentials," *Adv. Space Res.*, vol. 38, no. 11, pp. 2575–2580, 2006, doi: 10.1016/j.asr.2005.02.105.

[64] L. S. Matthews, K. Qiao, and T. W. Hyde, "Dynamics of a dust crystal with two different size dust species," *Adv. Space Res.*, vol. 38, no. 11, pp. 2564–2570, 2006, doi: 10.1016/j.asr.2005.06.037.

[65] K. Qiao, J. Kong, Z. Zhang, L. S. Matthews, and T. W. Hyde, "Mode Couplings and Conversions for Horizontal Dust Particle Pairs in Complex Plasmas," *IEEE Trans. Plasma Sci.*, vol. 41, no. 4, pp. 745–753, Apr. 2013, doi: 10.1109/TPS.2012.2236361.

[66] K. Qiao, J. Kong, E. V. Oeveren, L. S. Matthews, and T. W. Hyde, "Mode couplings and resonance instabilities in dust clusters," *Phys. Rev. E*, vol. 88, no. 4, p. 043103, Oct. 2013, doi: 10.1103/PhysRevE.88.043103.

[67] J. Elgeti, R. G. Winkler, and G. Gompper, "Physics of microswimmers—single particle motion and collective behavior: a review," *Rep. Prog. Phys.*, vol. 78, no. 5, p. 056601, Apr. 2015, doi: 10.1088/0034-4885/78/5/056601.

[68] C. Bechinger, R. Di Leonardo, H. Löwen, C. Reichhardt, G. Volpe, and G. Volpe, "Active particles in complex and crowded environments," *Rev. Mod. Phys.*, vol. 88, no. 4, p. 045006, Nov. 2016, doi: 10.1103/RevModPhys.88.045006.

[69] V. Nosenko, A. V. Ivlev, and G. E. Morfill, "Laser-induced rocket force on a microparticle in a complex (dusty) plasma," *Phys. Plasmas*, vol. 17, no. 12, p. 123705, Dec. 2010, doi: 10.1063/1.3525254.

[70] V. Nosenko, F. Luoni, A. Kaouk, M. Rubin-Zuzic, and H. Thomas, "Active Janus particles in a complex plasma," *Phys. Rev. Res.*, vol. 2, no. 3, p. 033226, Aug. 2020, doi: 10.1103/PhysRevResearch.2.033226.

[71] Y.-K. Tsang, "Nonuniversal velocity probability densities in two-dimensional turbulence: The effect of large-scale dissipation," *Phys. Fluids*, vol. 22, no. 11, p. 115102, Nov. 2010, doi: 10.1063/1.3504377.

[72] A. Bracco, J. LaCasce, C. Pasquero, and A. Provenzale, "The velocity distribution of barotropic turbulence," *Phys. Fluids*, vol. 12, no. 10, pp. 2478–2488, Sep. 2000, doi: 10.1063/1.1288517.

[73] C. Pasquero, A. Provenzale, and A. Babiano, "Parameterization of dispersion in two-dimensional turbulence," *J. Fluid Mech.*, vol. 439, pp. 279–303, Jul. 2001, doi: 10.1017/S0022112001004499.

[74] G. Livadiotis, "Chapter 5 - Basic Plasma Parameters Described by Kappa Distributions," in *Kappa Distributions*, G. Livadiotis, Ed. Elsevier, 2017, pp. 249–312. doi: 10.1016/B978-0-12-804638-8.00005-X.

[75] "Phys. Rev. E 48, 1683 (1993) - Self-similar transport in incomplete chaos." https://journals.aps.org/pre/abstract/10.1103/PhysRevE.48.1683 (accessed Sep. 14, 2018).

[76] T. S. Strickler, T. K. Langin, P. McQuillen, J. Daligault, and T. C. Killian, "Experimental Measurement of Self-Diffusion in a Strongly Coupled Plasma," *Phys. Rev. X*, vol. 6, no. 2, p. 021021, May 2016, doi: 10.1103/PhysRevX.6.021021.

[77] T. Ott, M. Bonitz, Z. Donkó, and P. Hartmann, "Superdiffusion in quasi-two-dimensional Yukawa liquids," *Phys. Rev. E*, vol. 78, no. 2, p. 026409, Aug. 2008, doi: 10.1103/PhysRevE.78.026409.

[78] B. Liu and J. Goree, "Superdiffusion and Non-Gaussian Statistics in a Driven-Dissipative 2D Dusty Plasma," *Phys. Rev. Lett.*, vol. 100, no. 5, p. 055003, Feb. 2008, doi: 10.1103/PhysRevLett.100.055003.

[79] B. Liu and J. Goree, "Superdiffusion in two-dimensional Yukawa liquids," *Phys. Rev. E*, vol. 75, no. 1, p. 016405, Jan. 2007, doi: 10.1103/PhysRevE.75.016405.

[80] L.-J. Hou, A. Piel, and P. K. Shukla, "Self-Diffusion in 2D Dusty-Plasma Liquids: Numerical-Simulation Results," *Phys. Rev. Lett.*, vol. 102, no. 8, p. 085002, Feb. 2009, doi: 10.1103/PhysRevLett.102.085002.

[81] S. Nunomura, D. Samsonov, S. Zhdanov, and G. Morfill, "Self-Diffusion in a Liquid Complex Plasma," *Phys. Rev. Lett.*, vol. 96, no. 1, p. 015003, Jan. 2006, doi: 10.1103/PhysRevLett.96.015003.

[82] T. Ott and M. Bonitz, "Anomalous and Fickian Diffusion in Two-Dimensional Dusty Plasmas," *Contrib. Plasma Phys.*, vol. 49, no. 10, pp. 760–764, Dec. 2009, doi: 10.1002/ctpp.200910089.

[83] O. S. Vaulina and S. V. Vladimirov, "Diffusion and dynamics of macro-particles in a complex plasma," *Phys. Plasmas*, vol. 9, no. 3, pp. 835–840, Feb. 2002, doi: 10.1063/1.1449888.

[84] T. Ott and M. Bonitz, "Is Diffusion Anomalous in Two-Dimensional Yukawa Liquids?," *Phys. Rev. Lett.*, vol. 103, no. 19, p. 195001, Nov. 2009, doi: 10.1103/PhysRevLett.103.195001.

[85] E. G. Kostadinova, J. L. Padgett, C. D. Liaw, L. S. Matthews, and T. W. Hyde, "Numerical study of anomalous diffusion of light in semicrystalline polymer structures," *Phys. Rev. Res.*, vol. 2, no. 4, p. 043375, Dec. 2020, doi: 10.1103/PhysRevResearch.2.043375.

[86] V. V. Uchaikin, "Self-similar anomalous diffusion and Levy-stable laws," *Phys.-Uspekhi*, vol. 46, no. 8, p. 821, 2003, doi: 10.1070/PU2003v046n08ABEH001324.

[87] R. Metzler and J. Klafter, "The random walk's guide to anomalous diffusion: a fractional dynamics approach," *Phys. Rep.*, vol. 339, no. 1, pp. 1–77, Dec. 2000, doi: 10.1016/S0370-1573(00)00070-3.

[88] T. Frugé Jones, E. G. Kostadinova, J. L. Padgett, and Q. Sheng, "A series representation of the discrete fractional Laplace operator of arbitrary order," *J. Math. Anal. Appl.*, vol. 504, no. 1, p. 125323, Dec. 2021, doi: 10.1016/j.jmaa.2021.125323.

[89] C. Liaw, "Approach to the Extended States Conjecture," *J Stat Phys*, vol. 153, pp. 1022–1038, Nov. 2013, doi: 10.1007/s10955-013-0879-5.

[90] D. Hundertmark, "A short introduction to Anderson localization," in *Analysis and stochastics of growth processes and interface models*, Oxford: Oxford University Press, 2008, pp. 194–218.

[91] H. Charan and R. Ganesh, "Molecular dynamics study of flow past an obstacle in strongly coupled Yukawa liquids," *Phys. Plasmas*, vol. 23, no. 12, p. 123703, Dec. 2016, doi: 10.1063/1.4971449.

[92] A. Douglass, V. Land, K. Qiao, L. Matthews, and T. Hyde, "Determination of the levitation limits of dust particles within the sheath in complex plasma experiments," *Phys. Plasmas 1994-Present*, vol. 19, no. 1, p. 013707, Jan. 2012, doi: 10.1063/1.3677360.

[93] V. Land, L. S. Matthews, T. W. Hyde, and D. Bolser, "Fluid modeling of void closure in microgravity noble gas complex plasmas," *Phys. Rev. E*, vol. 81, no. 5, p. 056402, May 2010, doi: 10.1103/PhysRevE.81.056402.

[94] J. D. Williams and E. Thomas, "Initial measurement of the kinetic dust temperature of a weakly coupled dusty plasma," *Phys. Plasmas*, vol. 13, no. 6, p. 063509, Jun. 2006, doi: 10.1063/1.2214640.
[95] J. D. Williams and E. Thomas, "Measurement of the kinetic dust temperature of a weakly coupled dusty plasma," *Phys. Plasmas*, vol. 14, no. 6, p. 063702, Jun. 2007, doi: 10.1063/1.2741457.
[96] B. Liu, J. Goree, M. Y. Pustylnik, H. M. Thomas, V. E. Fortov, A. M. Lipaev, A. D. Usachev, V. I. Molotkov, O. F. Petrov, and M. H. Thoma, "Particle velocity distribution in a three-dimensional dusty plasma under microgravity conditions," *AIP Conf. Proc.*, vol. 1925, no. 1, p. 020005, Jan. 2018, doi: 10.1063/1.5020393.

## Appendix A. MANY-BODY SIMULATION OF DUSTY PLASMA

In Sec. II B, we briefly outlined some features of the many-body simulations, which were most relevant to the discussion. Here we elaborate on each term in the force equation (repeated for convenience), which was used to obtain the dust particle dynamics

$$m_d \ddot{x} = \mathbf{F}_{dd} + m_d \mathbf{g} + Q_d \mathbf{E} - \beta \dot{x} + \zeta \mathbf{r}(t). \quad (A1)$$

Below we discuss the dust-dust interaction force $\mathbf{F}_{dd}$, the interplay between gravity $m_d \mathbf{g}$ and confinement forces $Q_d \mathbf{E}$, the drag force $\beta \dot{x}$, and the thermal bath $\zeta \mathbf{r}(t)$ used in the many-body simulations of the present study.

1. Interparticle potential

In typical complex plasma experiments, the primary dust-charging mechanism is the collection of electrons and ions from the environment. Due to the electrons' higher thermal speed, dust grains generally become negatively charged and surrounded by a region of positive space charge due to ion shielding. The resulting *local* interparticle interaction is commonly assumed to be of the Yukawa (screened Coulomb) form

$$V(r) = \left(\frac{Q_d}{4\pi\varepsilon_0 r}\right) e^{-\frac{r}{\lambda_D}}, \quad (A2)$$

where $r$ is the distance from a particle with charge $Q_d$ and $\lambda_D$ is the Debye shielding length (the scale length over which a charged grain is shielded by the plasma). It has been reported [91] that in strongly coupled liquids with Yukawa interactions, the laminar-to-turbulent transition at a fixed Reynolds number $Re$ can be induced by increasing the strength of the interparticle potential alone. This behavior has been verified across different ranges of Reynolds numbers (low $Re < 0.1$, intermediate $[2 \leq Re \leq 35]$, and high $Re > 50$). Thus, in the present study of active turbulence, it was considered appropriate to derive the dust-dust interaction force $F_{dd}$ from a Yukawa potential. In a future study, it will be interesting to consider how modifications on the Debye spheres and the corresponding dust-dust interaction influence the observed dust diffusion and the onset of turbulent instabilities.

2. Confinement forces and gravity

The confinement forces in equation (1) / (A1) are electrostatic in nature and have the form

$$\mathbf{F}_{conf} = m_d \mathbf{a}_{conf} = Q_d \mathbf{E} \Rightarrow \mathbf{a} = \frac{Q_d}{m_d} \mathbf{E}, \quad (A3)$$

where the radial confinement is derived from the 10$^{\text{th}}$ order polynomial potential

$$V_i = 0.5 m_d \left( \frac{\Omega_h^2 \rho_i^{10}}{R^8} \right), \tag{A4}$$

which was discussed in Sec. II B. The vertical confinement in a dusty plasma experiment is typically provided by the electrodes in the cell. Here we model such confinement by a 5th order polynomial electrostatic field which was obtained from a fluid model of the plasma in a GEC cell [92], [93]

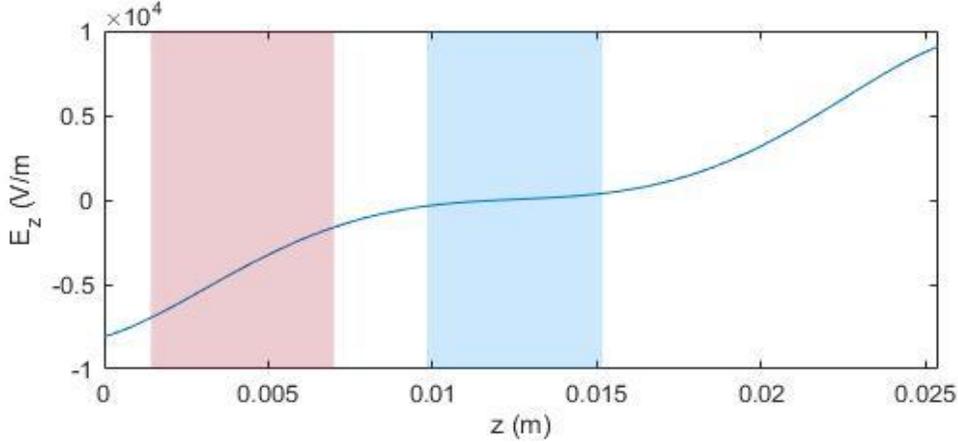

Fig. 9. Fifth order polynomial used for the vertical electric field. The red shaded region is the sheath above the lower electrode, where the electric field is approximately linear. The region shaded blue is the plasma bulk, where the electric field is essentially zero. The region between the bulk and the sheath is the pre-sheath.

$$\boldsymbol{E}_{z,conf}(z) = [-8083 + 553373 z + (2.0 \times 10^8) z^2 - (3.017 \times 10^{10}) z^3 \\ + (1.471 \times 10^{12}) z^4 - (2.306 \times 10^{13}) z^5] \hat{\boldsymbol{z}}, \tag{A5}$$

where z is the height above the lower electrode, as shown in Figure 9.

Note that $\hat{\boldsymbol{z}}$ is positive in the direction from the lower to the upper electrode, which is opposite to the direction of gravity. In the sheath region of the plasma, the direction of the electric field coincides with the direction of gravity, i.e., $(-\hat{\boldsymbol{z}})$. Since the dust particles are negatively charged, the resulting electric field force is in the $(+\hat{\boldsymbol{z}})$ direction, balancing the force due to gravity, which allows dust levitation at the exact location where forces balance. The 5th order polynomial was obtained from fits to a fluid model code and experimental data, discussed in [53]. This polynomial fit was obtained for particular plasma conditions. In general, as the power delivered to the plasma increases, the confinement force becomes stronger with a steeper $E_z$. In the region where the dust is levitating in the sheath, the electric field is often approximated as linear. The bottom electrode is assumed to be located at $z = 0 m$, and the top at $z = 0.0254 m$, after which the electric field takes a constant value of $8083 \, V/m$.

As a result, a dust grain of mass m and negative charge of magnitude $Q_d$ acquires acceleration

$$\boldsymbol{a}_{z,conf}(z) = -\frac{Q_d}{m_d} \boldsymbol{E}_{z,conf}(z) = -\frac{Q_d}{m_d} E_{z,conf}(z)(-\hat{\boldsymbol{z}}) = \frac{Q_d}{m_d} E_{z,conf}(z) \hat{\boldsymbol{z}}. \tag{A6}$$

This electric force is exerted against gravity, so that the electro-gravitational (eg) acceleration in the z direction is given in the simulation by

$$\boldsymbol{a}_{z,eg}(z) = \frac{Q_d}{m_d} E_{z,conf}(z)\hat{\boldsymbol{z}} - 9.81\hat{\boldsymbol{z}}. \tag{A7}$$

3. Gas drag and thermal bath

The gas drag force in the many-body simulation is given by $\boldsymbol{F}_{drag} = -\beta\dot{\boldsymbol{x}}$, where the damping factor $\beta$ depends on the neutral gas pressure and temperature, with

$$\beta = \delta \frac{4\pi}{3}(r_d)^2 n \frac{m_g}{m_d} \sqrt{\frac{8k_B T_g}{\pi m_g}}. \tag{A8}$$

Here $\delta = 1.44$ (measured for melamine formaldehyde dust in argon), $r_d$ is the dust radius, $n$ the gas number density, $m_g$ the molecular mass of the gas, $T_g$ the gas temperature (assumed to be 300 K), and $m_d$ is the mass of the dust. The Epstein damping times for the simulations presented here are 0.1497 s (at 5 Pa), 0.7484 s (at 1 Pa) and 7.4838 s (at 0.1 Pa). Note that even for the longest damping time, $\approx 7.5\ s$, the 30 s simulation time covers 4 damping periods.

A thermal bath is realized by subjecting the particles to random force "kicks", $F_{th} = \zeta \boldsymbol{r}(t)$ with the maximum acceleration imparted by the amplitude

$$\zeta = \sqrt{\frac{2\beta k_B T_g}{m_d \Delta t_d}}, \tag{A9}$$

where $\Delta t_d$ is the dust time step. Notice that each kick represents a cumulative effect from neutral collisions with the dust particle, which yield a measurable displacement at the characteristic timescale of the dust. The random number $\boldsymbol{r}(t)$ is selected from a Gaussian distribution with a unit normal distribution (zero mean and unit variance). Gaussian distribution of random variables should yield a classical diffusion of the dust grains (also called, Brownian motion or a random walk). The dust particles are expected to exhibit classical diffusion in cases where all interactions are local, which is true for collisions with neutral gas particles from the environment. A Gaussian distribution with a narrow width leads to small random numbers $\boldsymbol{r}(t)$, which is appropriate for modeling strongly coupled dust crystals at high gas pressure. As the variance of the Gaussian is increased, bigger random numbers $\boldsymbol{r}(t)$ are more likely, which leads to less stable dust grains. Thus, in the present study, the thermal bath is responsible for mobilization of dust grains at lower pressures. However, the observed anomalous dust diffusion results from nonlocal interactions due to charge effects, and not from the thermal bath.

## Appendix B. DUST TEMPERATURE AND VELOCITY DISTRIBUTIONS

Extracting temperature from Gaussian fits is not straightforward for interacting particles in general, and is particularly difficult for dusty plasma monolayers, which are known to exhibit non-Maxwellian velocity PDFs. Additionally, the kinetic temperature measured from fits to the velocity histograms is known to be highly anisotropic and larger than room temperature at low pressures (as discussed in [94], [95]). In Fig. 10, we show different fits to the velocity PDFs for each pressure case. For each pressure, the $v_z$ components exhibit strongly non-Maxwellian distributions and are best described by a combination of Maxwellian distribution and Cauchy distribution tails. The $v_x$ and $v_y$ components in each case can be described by a Maxwellian fits, which indicate that the temperature increases substantially as pressure decreases: $\approx 130 - 135\ K$ (at 5 Pa), $\approx 650 - 700\ K$ (at 1 Pa), and $\approx 4300\ K$ (at 0.1 Pa). Although the kinetic temperature extracted for the 0.1 Pa case seems very high, temperatures on the order of eV are consistent with temperatures reported by Williams and Thomas in [94], [95]

Alternatively, a one can fit a Maxwellian distribution with Kappa tails to the $v_x$ and $v_y$ PDFs in the 0.1 Pa case, which yields a much lower temperature $\approx 400\ K$. This is consistent with the temperature obtained from the fit to the $v_z$ PDF. The parameter $\kappa$ characterizes how far the system is from thermal equilibrium. When $\kappa \to \infty$, the Kappa distribution approaches a Maxwellian, but when $\kappa$ is finite, the distribution function has high-energy tails, with more high velocity particles than for a Maxwellian. For $\kappa < 10$, the Kappa distribution has a power law tail. Here, $\kappa \approx 11$ for both $v_x$ and $v_y$ PDFs, suggesting that the 0.1 Pa case is away from thermal equilibrium but does not exhibit power-law tails. The modified Maxwellian distribution with Kappa tails has been used to describe the velocity PDFs of weakly coupled 3D dusty plasma cloud, suspended in microgravity Neon plasma, powered by a RF coil [96]. Those studies found $\kappa = 1.72$ and $T = 161\ K$. Note that these experiments did not include external heating or perturbation.

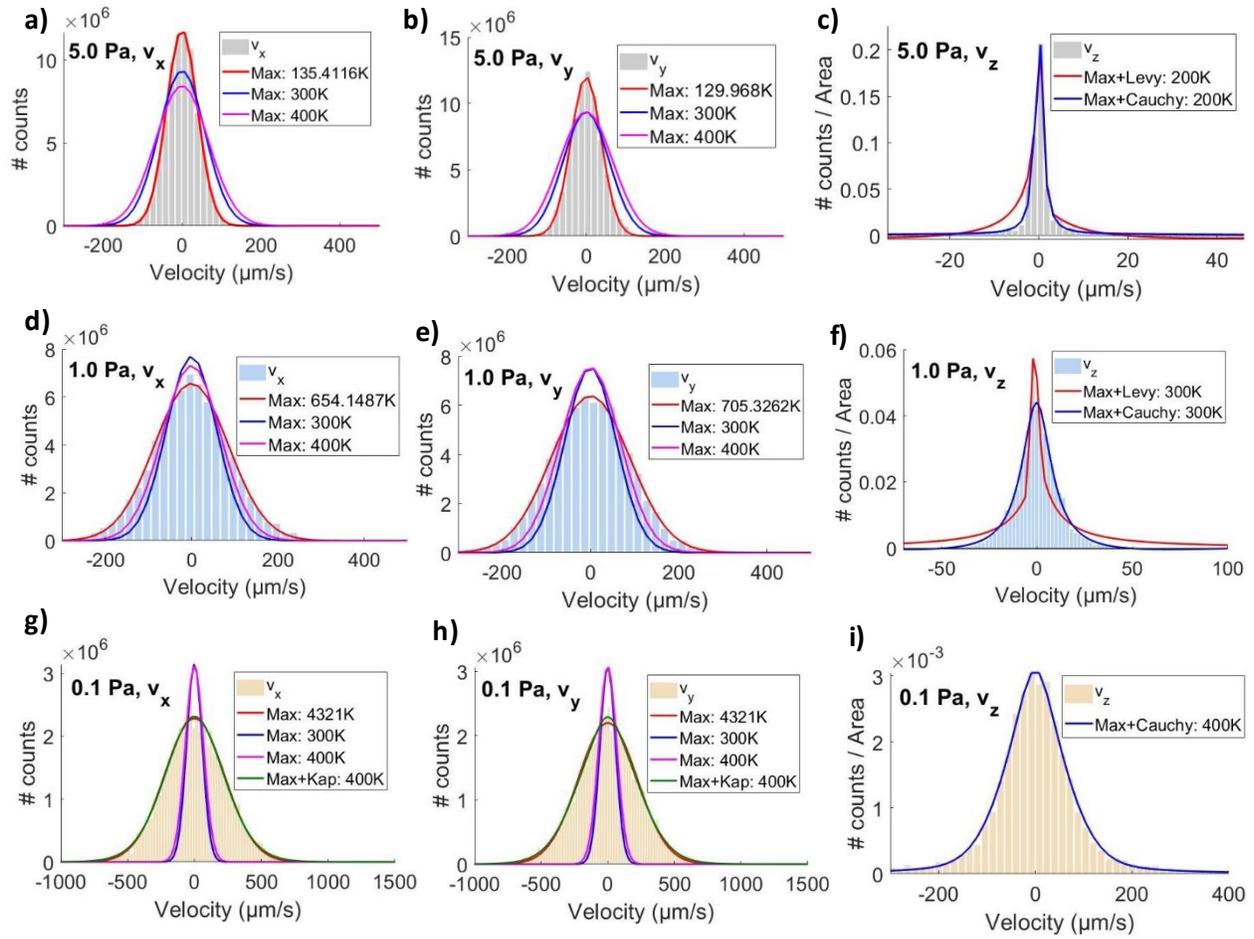

Fig. 10. Plots of the velocity histograms for a), b), c) 5 $Pa$, d), e), f) 1 $Pa$, and g), h), i) 0.1 $Pa$. In each case, the velocities were extracted from 10000 particles tracked for 10 s at 1000 Hz output frequency, yielding $10^8$ data points for the statistics.